\documentclass[review]{elsarticle}

\usepackage{lineno}
\usepackage{hyperref}
\modulolinenumbers[5]

\journal{NeuroImage}

\usepackage{amsmath}
\usepackage{amssymb}
\usepackage{amsthm}
\usepackage{mathtools}
\usepackage{graphicx}
\usepackage{microtype}
\usepackage{color}
\usepackage{dsfont}
\usepackage{booktabs}
\usepackage{tikz}

\usetikzlibrary{arrows}

\graphicspath{ {./figures/} }

\DeclareMathOperator*{\argmax}{arg\,max}
\DeclareMathOperator*{\argmin}{arg\,min}

\newtheorem{proposition}{Proposition}

\newcommand{\referencesToColor}{}

\newcommand{\beginsupplement}{%
        \setcounter{table}{0}
        \renewcommand{\thetable}{S\arabic{table}}%
        \setcounter{figure}{0}
        
        \renewcommand{\figurename}{Inline Supplementary Figure}
        \renewcommand{\thefigure}{\arabic{figure}}%
     }
     
\biboptions{authoryear}

\begin{document}

\begin{frontmatter}

\title{
A multivariate nonlinear mixed effects model for longitudinal image analysis:
Application to amyloid imaging\tnoteref{t1,t2}}

\tnotetext[t1]{DOI: \href{dx.doi.org/10.1016/j.neuroimage.2016.04.001}{10.1016/j.neuroimage.2016.04.001}}
\tnotetext[t2]{ This manuscript version is made available under the CC-BY-NC-ND~4.0 license
\url{http://creativecommons.org/licenses/by-nc-nd/4.0/}.}

\author[jhu-iacl,jhu-bme,nia-lbn]{Murat~Bilgel\corref{corresponding-author}}
\cortext[corresponding-author]{Corresponding author at: 
National Institute on Aging,
Laboratory of Behavioral Neuroscience,
251 Bayview Blvd.,
Suite 100, Rm 04B316,
Baltimore, MD 21224, USA.
Phone: +1-410-558-8151.
Fax: +1-410-558-8674.}
\ead{murat.bilgel@nih.gov}

\author[jhu-iacl,jhu-bme,jhu-ece,jhu-rad]{Jerry~L.~Prince}
\author[jhu-rad]{Dean~F.~Wong}
\author[nia-lbn]{Susan~M.~Resnick}
\author[psu-math]{Bruno~M.~Jedynak}

\address[jhu-iacl]{Image Analysis and Communications Laboratory, Johns Hopkins University School of Engineering, Baltimore, MD, USA}
\address[jhu-bme]{Dept. of Biomedical Engineering, Johns Hopkins University School of Medicine, Baltimore, MD, USA}
\address[nia-lbn]{Laboratory of Behavioral Neuroscience, National Institute on Aging, National Institutes of Health, Baltimore, MD, USA}
\address[jhu-ece]{Dept. of Electrical and Computer Engineering, Johns Hopkins University School of Engineering, Baltimore, MD, USA}
\address[jhu-rad]{Dept. of Radiology, Johns Hopkins University School of Medicine, Baltimore, MD, USA}
\address[psu-math]{Dept. of Mathematics and Statistics, Portland State University, Portland, OR, USA}

\begin{abstract}
  It is important to characterize the temporal trajectories of disease-related 
  biomarkers in order to monitor progression and identify potential points of intervention.
  These are especially important for neurodegenerative diseases, as therapeutic 
  intervention is most likely to be effective in the preclinical disease stages
  prior to significant neuronal damage.
  Neuroimaging allows for the measurement of structural, functional, and 
  metabolic
  integrity of the brain at the level of voxels, whose volumes are on the order of 
  mm$^3$.
  These voxelwise measurements provide a rich collection of disease indicators.
  Longitudinal neuroimaging studies enable the analysis of  
  changes in these voxelwise measures.
  However, commonly used longitudinal analysis approaches, such as linear mixed 
  effects models, do not account for the fact that
  individuals enter a study at various disease stages and progress at different 
  rates, and generally consider each voxelwise measure independently.
  We propose a multivariate nonlinear mixed effects model for estimating the trajectories
  of voxelwise neuroimaging biomarkers from longitudinal data that accounts for
  such differences across individuals.
  The method involves the prediction of a progression score for each visit 
  based on a collective analysis of voxelwise biomarker data within an
  expectation-maximization framework 
  that efficiently handles large amounts of measurements and variable number of 
  visits per individual,
  and accounts for spatial correlations among voxels.
  This score allows individuals with similar progressions to be aligned and 
  analyzed together, which enables the construction of a trajectory of brain changes as a function of
  an underlying progression or disease stage.
  We apply our method to studying cortical $\beta$-amyloid deposition,
  a hallmark of preclinical Alzheimer's disease, as measured using 
  positron emission tomography.
  Results on 104 individuals with a total of 300 visits
  suggest that precuneus is the 
  earliest cortical region to accumulate amyloid, 
  closely followed by the cingulate 
  and frontal cortices, then by the lateral parietal cortex.
  The extracted progression scores reveal a pattern similar to mean cortical 
  distribution volume ratio~(DVR), an index of global brain amyloid levels.
  The proposed method can be applied to other types of longitudinal imaging data, including
  metabolism, blood flow, tau, and structural imaging-derived measures,
  to extract individualized summary scores indicating disease progression and to 
  provide voxelwise trajectories that can be compared between brain regions.
\end{abstract}


\end{frontmatter}
\linenumbers

\section{Introduction}
It is important to characterize the temporal trajectories of disease-related 
biomarkers in order to monitor progression and to identify potential points of intervention.
Such a characterization is
especially important for neurodegenerative diseases, as therapeutic 
intervention is most likely to be effective in the preclinical disease stages
prior to significant neuronal damage.
For example, in Alzheimer's disease, brain changes evident in structural, 
functional, and metabolic imaging may occur more than a decade before the onset of 
cognitive symptoms~\citep{Bateman2012}, with cortical amyloid-$\beta$~(A$\beta$) accumulation
being one of the earliest changes~\citep{Jack2013,Sperling2014b,Villemagne2013}.
Such brain changes can be measured using neuroimaging techniques and can be tracked 
over time at the individual level via longitudinal studies.

Given the focus on preventing and delaying the onset of incurable neurodegenerative
diseases, the emphasis of 
clinical trials has shifted to studying clinically normal individuals with 
positive biomarkers, for example those exhibiting brain amyloid in the case of AD,
in order to identify early intervention opportunities in the 
preclinical stages of disease~\citep{Sperling2014}.
It is important to determine the temporal trajectories of hypothesized 
biomarkers in the early 
disease stages in order to better understand their associations with disease 
progression.
Current neuroimaging methods allow for the characterization of the brain at the mm$^3$ 
level, generating hundreds of thousands of measurements that can be used 
as potential biomarkers of neurodegenerative diseases.
Understanding the temporal trajectories of these voxelwise measurements can 
provide clues into disease mechanisms by identifying the earliest and fastest changing
brain regions.

Changes in voxelwise neuroimaging measurements over time are commonly studied using
linear mixed effects models~\citep{Bernal-Rusiel2012,Bernal-Rusiel2013,Ziegler2015}.
Univariate linear mixed effects models use time or age to characterize changes in a single
imaging measure. 
However, time or age may not be the appropriate metric for measuring disease progression 
due to variability across individuals. 
While covariates can be included in linear mixed effects models to account for this variability, 
choosing the correct set of covariates is difficult and covariates generally have a more 
complicated association with disease progression than the assumed linear relationship of 
linear mixed effects models. Instead, this variability can be accounted for by aligning individuals
in time based on their longitudinal biomarker profiles within a multivariate framework.
This is the premise of the Disease Progression Score method, 
which has been applied to studying changes in cognitive and biological markers
related to Alzheimer's disease \citep{Jedynak2012,Jedynak2014,Bilgel2014b}.
It is assumed that there is an underlying 
progression score~(PS) for each subject visit that is an affine transform of the 
subject's age, and given this PS, it is possible to place biomarker 
measurements across a group of subjects onto a common timeline.
The affine transformation of age removes across-subject variability in baseline 
biomarker measures as well as in their rates of longitudinal progression.
Each biomarker is associated with a parametric trajectory as a function of PS,
whose parameters are estimated along with the PS for each subject.
This allows one to ``stitch" data across subjects to obtain temporal 
biomarker trajectories that fit an underlying model~(Fig.~\ref{fig:DPS-illustration}).

Previous approaches have used certain cognitive measures, such as
ADAS-Cog~\citep{Caroli2010,Yang2011},
MMSE~\citep{Doody2010} or CDR-SB~\citep{Delor2013} as a surrogate for disease 
progression to delineate the trajectories of other AD-related cognitive 
measurements.
These methods operate with the assumption that disease progression is reflected
by a single cognitive measurement rather than a profile of multiple 
measurements, and therefore are inherently limited in their characterization of disease 
evolution.
\citet{Younes2014} fitted a piecewise linear model to longitudinal data
assuming that each biomarker becomes abnormal a certain 
number of years before clinical diagnosis, and
this duration was estimated for each biomarker to yield
longitudinal trajectories as a function of time to diagnosis.
A quantile regression approach was employed by \citet{Schmidt-Richberg2015} to align a 
sample of cognitively normals and mild cognitively impaired~(MCI) with a sample 
of MCI and AD, and then to estimate biomarker trajectories.
These approaches assume that all individuals are on a path to disease and
require knowledge of clinical diagnosis.
Therefore, they are not 
suitable for studying the earliest changes in individuals who have not converted
to a clinical diagnosis.
\citet{Donohue2014} applied a self-modeling regression model
within a multivariate framework to characterize
the longitudinal trajectories of a set of cognitive, CSF, and neuroimaging-based 
biomarkers.
This approach allows for across-subject variability only in the age of onset, not
in progression speed.
Models incorporating fixed effects as well as individual-level random 
effects have been proposed to study
ADAS-Cog~\citep{Ito2011,Schiratti2015} and regional cortical 
atrophy~\citep{Schiratti2015},
and \citet{Schulam2015} used a spline model that incorporates longitudinal 
clustering and modeling of individual-level effects to study trajectories of
scleroderma markers.
These mixed effects models take into consideration each measure separately 
rather than using them within a unifying framework.
Others have used event-based probabilistic frameworks to 
determine the ordering of changes in longitudinal biomarker
measures as well as the appropriate
thresholds for separating normal from abnormal measures~\citep{Fonteijn2012,Young2014}.
These methods
characterize longitudinal biomarker trajectories in a discrete framework
rather than a continuous one.
\citet{Schiratti2015a} proposed an extension to their earlier approach to 
model multiple measures together.
Biomarker trajectories are assumed to be identical except for a 
shift along the disease timeline, and this assumption
prevents hypothesis testing regarding rate of change across biomarkers.
Furthermore, biomarkers are assumed to be conditionally independent given the 
subject-level random effects, but this assumption is not realistic when biomarkers 
are voxel-based neuroimaging measurements.

Here, we adapt the disease progression score principle to studying longitudinal 
neuroimaging data by making substantial innovations to the progression score model and
parameter estimation procedure. 
First, voxelwise imaging measures
constitute the biomarkers in the model, and are analyzed together in a 
multivariate framework.
Studying progression at the voxel level rather than using region of 
interest~(ROI)-based measures 
allows for the discovery of patterns that may not 
be confined within any given ROI.
Second, since voxelwise imaging measures have an underlying spatial correlation, 
we incorporate the modeling of the spatial correlations among 
the biomarker error terms.
Modeling spatial correlations makes
the inference of the subject-specific progression scores less 
susceptible to the inherent correlations among the voxels.
Third, we incorporate a bivariate normal prior on the subject-specific variables 
that define the relationship between age and PS.
The prior allows a better modeling of the variance within individuals and
enables the incorporation of individuals with a single visit into the 
model fitting procedure.
Fourth, instead of using an alternating least-squares approach
for parameter estimation as presented by~\citet{Jedynak2012},
we formulate the model fitting as an expectation-maximization~(EM) 
algorithm, which guarantees convergence to a local maximum
and allows for an efficient model fitting framework for a large number of biomarkers.
Finally, we present a statistical framework for comparing 
the onset and rate of progression across different regions.
This paper extends our previous approach for analyzing longitudinal voxelwise 
imaging measures using the progression score framework by incorporating
a prior on the subject-specific variables, presenting a hypothesis testing 
framework for determining biomarker ordering, and performing an extensive
validation of the method~\citep{Bilgel2015a}.

We first show using simulated data that the model parameters are estimated accurately
and that modeling spatial correlations improves parameter estimation. 
We then apply the method to distribution 
volume ratio~(DVR) images derived from Pittsburgh compound~B~(PiB) PET imaging, 
which show the distribution of cerebral fibrillar amyloid.
Models fitted using data for 104 participants with a total of 300 PiB-PET visits
reveal that the precuneus and frontal cortex show the greatest longitudinal increases in
fibrillar amyloid,
with smaller increases in lateral temporal and temporoparietal regions, and minimal 
increases in the occipital cortex and the sensorimotor strip.
Our results suggest that the precuneus is
the earliest cortical region to accumulate amyloid.
The results are consistent across the two hemispheres, and the estimated PS agrees with
a widely used PET-based global index of brain amyloid known as mean cortical DVR.
The presented method can be applied to other types of longitudinal imaging data 
to understand voxelwise trajectories and to quantify each individual scan 
against the estimated progression pattern.

\section{Method}

\subsection{Model}
Our goal is to characterize the progression of disease or an underlying process
as measured using a collection of relevant biomarkers.
Disease or process stage, as indicated by a progression score~(PS), $s$,
is intrinsically related to time $t$, measured as the age of a subject.
Since individuals differ in their onset and rate of progression,
the relationship between $s$ and $t$ varies across individuals.
We model the progression $s$ as a linear function of time $t$ for each individual
and allow for the prediction of separate slopes and intercepts to account for 
this variability across individuals.

\begin{figure}
  \centering
  \includegraphics[width=1\textwidth]{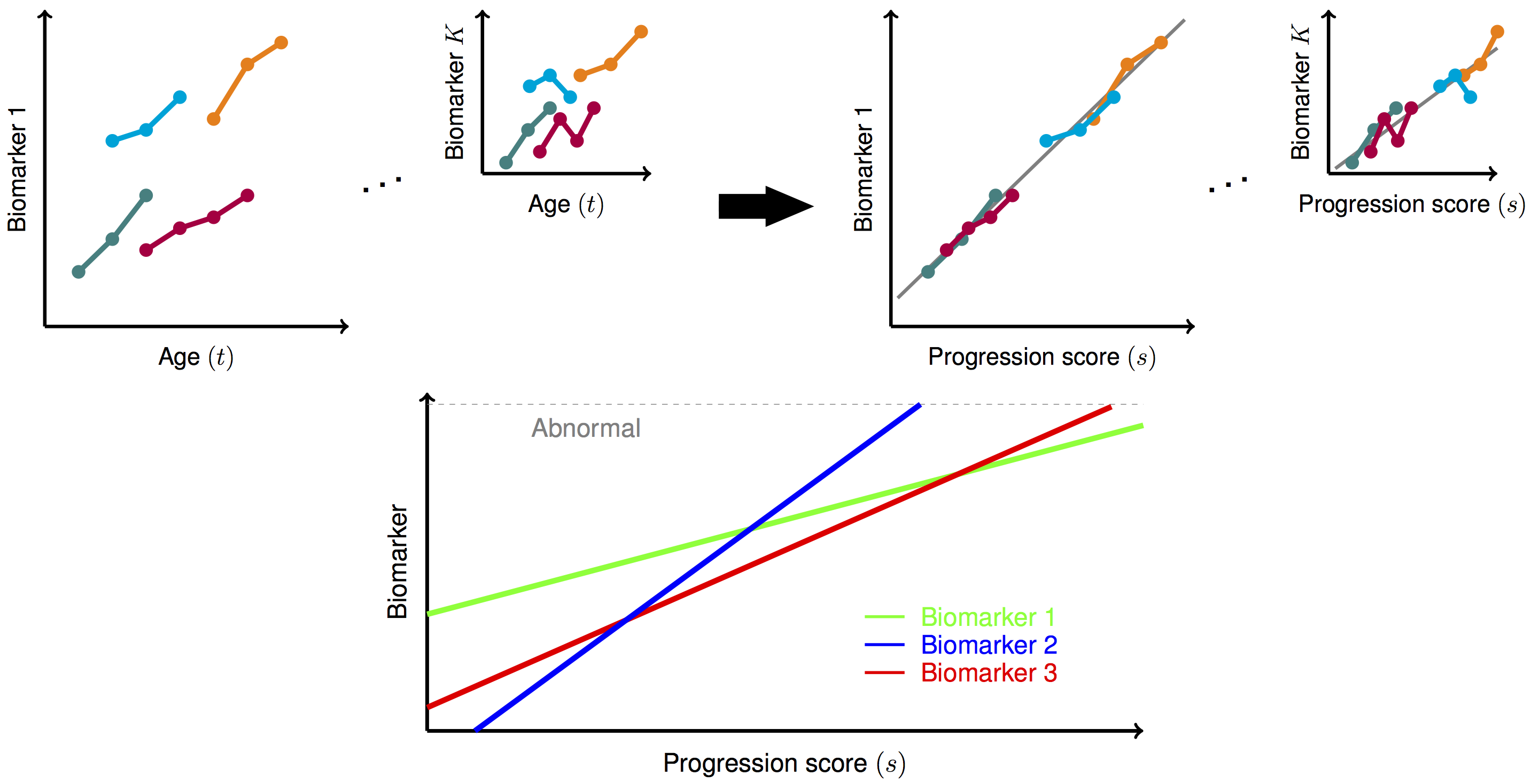}
  \caption{Illustration of the biomarker alignment concept in the progression 
  score model.
  The biomarkers we consider in this work are PET 
  measures of cerebral amyloid across a total of $K \approx 30,000$ voxels.
  \emph{Top:} Progression score~(PS) aligns longitudinal measures better than age,
  and allows for the estimation of a trajectory for each
  biomarker/voxel (in gray).
  \emph{Bottom:} Estimated biomarker trajectories can be compared on the common PS scale.
  \referencesToColor}
  \label{fig:DPS-illustration}
\end{figure}

Generally, there is a particular presentation of symptoms and biomarker 
measurements at a given progression stage.
Furthermore, as the disease or process progresses, there is a particular 
temporal progression of the biomarkers.
In this work, we
consider voxelwise PET measures as biomarkers and
model the temporal trajectory of each biomarker, or voxel,
as a linear function of the progression $s$.
Considering voxelwise neuroimaging measures as biomarkers may appear unusual;
however, these measures fit the NIH definition of a biomarker: ``a characteristic that is
objectively measured and evaluated as an indicator of normal biological processes,
pathogenic processes, or pharmacologic responses to a therapeutic
intervention''~\citep{Definitions2001}.
As illustrated in Figure~\ref{fig:DPS-illustration}, PS aligns longitudinal biomarker 
measures better than age since it accounts for differences across individuals in 
rates as well as baseline levels of progression.
After this alignment in time,
the estimated biomarker trajectories can be compared on the common PS 
scale.

In the following subsections, we describe the progression score model in detail.

\subsubsection{Subject-specific model}
The progression score $s_{ij}$ for subject $i$ at visit $j$ is assumed to be an
affine transformation of the subject's age $t_{ij}$:
\begin{eqnarray}
  s_{ij} &=& \alpha_{i} t_{ij} + \beta_i \nonumber \\
  &=& \mathbf{q}_{ij}^T \mathbf{u}_i ,
\end{eqnarray}
where
$\mathbf{q}_{ij} = 
\begin{bmatrix}
  t_{ij} \\ 1
\end{bmatrix}$, and
$ \mathbf{u}_i =
\begin{bmatrix}
    \alpha_i \\
    \beta_i
  \end{bmatrix}
 $. The subject-specific variables,
 $\alpha_i$ and $\beta_i$ contained in the vector $\mathbf{u}_i$,
 are assumed to follow a bivariate normal distribution, i.e.,
$\mathbf{u}_i \sim \mathcal{N}_2(\mathbf{m},V)$,
which are independent and identically distributed across subjects.
This model accounts for differences between subjects in the rate of progression via 
$\alpha$, and in the baseline levels of disease progression via $\beta$.

\subsubsection{Subject-specific prior covariance model}
\label{sec:prior-cov-model}
The prior covariance $V$ is modeled as a $2 \times 2$ unstructured covariance matrix.
Log-Cholesky parametrization of $V$, given by $\boldsymbol \nu$,
ensures that $V \equiv V(\boldsymbol \nu)$ is positive definite~\citep{Pinheiro1996}.
Let $U = 
\begin{bmatrix}
  U_{11} & U_{12} \\ 0 & U_{22}
\end{bmatrix}$ be an upper triangular matrix such that $V = U^T U$.
If the diagonal elements of $U$ are constrained to be positive, then this 
Cholesky decomposition is unique.
To ensure that the diagonal elements of $U$ are positive in an unconstrained optimization
framework, we use the natural logarithm of the diagonal elements of $U$ as parameters.
We then vectorize the upper triangular elements (including the diagonal) of $U$ to obtain the 
parameter vector 
$\boldsymbol \nu = [\log U_{11} \;\; U_{12} \;\; \log U_{22}]^T$, 
which uniquely parameterizes $V$ and ensures its positive definiteness.

\subsubsection{Biomarker trajectory model}
The collection of $K$ biomarker measurements form the $K \times 1$ vector 
$\mathbf{y}_{ij}$ for subject $i$ at visit $j$.
Longitudinal trajectories associated with these biomarkers are assumed to be 
linear and
parameterized by $K \times 1$ vectors $\mathbf{a}$ and $\mathbf{b}$:
\begin{equation}
  \mathbf{y}_{ij} = \mathbf{a} \, s_{ij} + \mathbf{b} + \boldsymbol\epsilon_{ij} .
  \label{eq:model}
\end{equation}
Here, $\mathbf{a} = [a_1, a_2, \ldots, a_K]^T$,
$\mathbf{b} = [b_1, b_2, \ldots, b_K]^T$, and
$\boldsymbol\epsilon_{ij} \sim \mathcal{N}_K(0,R)$
is the observation noise.
$\boldsymbol\epsilon_{ij}$~are assumed to be independent and identically distributed across 
subjects and visits.

\subsubsection{Noise covariance model}
The matrix $R$ is assumed to have the form
$R = \Lambda C \Lambda$, where 
$\Lambda$ is a diagonal matrix with positive diagonal elements
$\boldsymbol \lambda$
and $C$ is a correlation matrix parameterized by $\boldsymbol \rho$.
This parameterization guarantees that $R$ is a positive definite 
matrix~\citep{Galecki2013-ch10}.
For ease of notation, we let
$\Lambda \equiv \Lambda ( \boldsymbol \lambda )$,
$C \equiv C(\boldsymbol \rho)$, and
$R \equiv R(\boldsymbol \lambda,\boldsymbol \rho)$.

When the biomarkers under consideration have a spatial organization, i.e., if they 
are voxelwise measurements from medical images, then the correlation matrix $C$
can be described as a function of the spatial distance $d \equiv d(k,k')$ between pairs of 
voxels indexed by $k$ and $k'$ as well as the spatial 
correlation parameter $\boldsymbol \rho$.
Possible univariate parameterizations (i.e., $\boldsymbol \rho = \rho \in \mathbb{R}$)
of $C$ are presented in Table~\ref{tab:corr-fun}.
All of these spatial correlation functions ensure that $C$ is a valid 
correlation matrix~\citep{Galecki2013-ch10}.
\begin{table}[h!]
\centering
\caption{Spatial correlation functions.}
\setlength{\tabcolsep}{10pt}
\begin{tabular}{rl}
  \toprule
  Exponential & $C_{kk'} = e^{-d/\rho}$ \\
  Gaussian & $C_{kk'} = e^{-(d/\rho)^2}$\\
  Rational quadratic & $C_{kk'} = \frac{1}{1+(d/\rho)^2}$\\
  Spherical & $C_{kk'} = \left(1-\frac{3}{2}\frac{d}{\rho} + \frac{1}{2}\left(\frac{d}{\rho}\right)^3 \right) \mathds{1}(d<\rho)$\\
  \bottomrule
\end{tabular}

\vspace{.1em}
$d \equiv d(k,k')$ is the spatial distance between
voxels indexed by $k$ and $k'$.
\label{tab:corr-fun}
\end{table}

\subsubsection{Overall model}
\label{sec:overall-model}
\begin{figure}
  \centering
  \includegraphics[width=.5\textwidth]{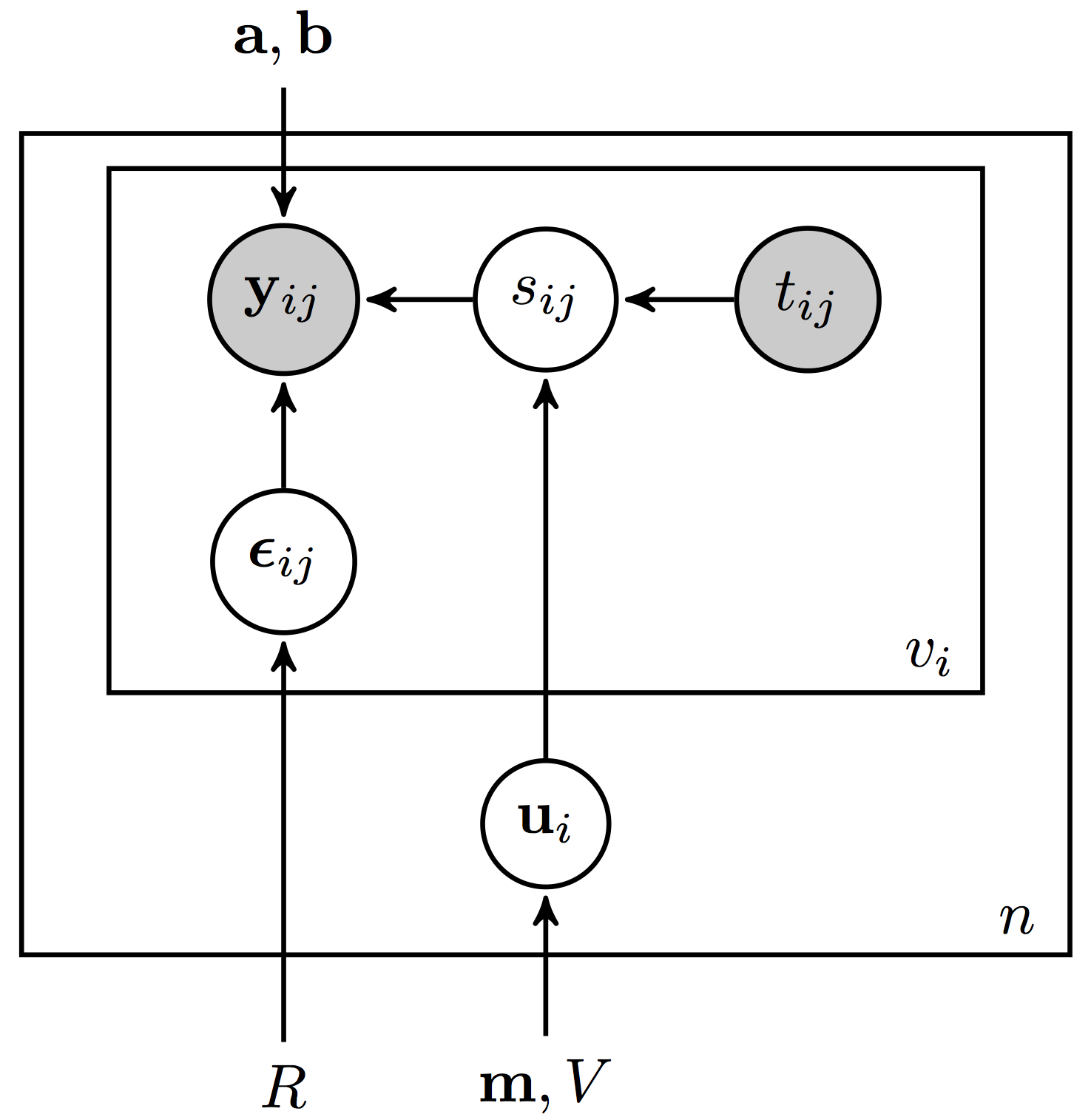}
  \caption{Probabilistic model of the progression score using plate notation.
  The progression score $s_{ij}$ establishes the link between age $t_{ij}$ and the voxelwise
  observations $\mathbf{y}_{ij}$. Circles indicate variables, and known measures are shaded.
  Rectangles indicate that the same model applies across visits (inner rectangle) and individuals
  (outer rectangle). Arrows indicate dependencies between variables and parameters.}
  \label{fig:model-diagram}
\end{figure}

The overall model, diagrammatically summarized using plate notation 
in Fig.~\ref{fig:model-diagram},
is described by the following equations:
\begin{eqnarray}
  s_{ij} & = & \mathbf{q}^T_{ij} \mathbf{u}_i \\
  \mathbf{y}_{ij} & = & \mathbf{a} \, s_{ij} + \mathbf{b} + \boldsymbol\epsilon_{ij} \\
  \mathbf{u}_i & \sim & \mathcal{N}_2(\mathbf{m},V(\boldsymbol\nu)) \\
  \boldsymbol\epsilon_{ij} & \sim & \mathcal{N}_K(0,R(\boldsymbol \lambda, \boldsymbol \rho)) 
  .
\end{eqnarray}

While this model is a mixed effects model since it incorporates
the fixed effects $\mathbf{a},\mathbf{b}$ as well as 
the individual-level random effects $\mathbf{u}_i$, and is nonlinear in the 
parameters, it departs from the form of the nonlinear mixed effects model 
described by~\citet{Lindstrom1990}.
Therefore, instead of pursuing a restricted maximum likelihood approach, we use 
an expectation-maximization~(EM) approach, as described below.

Let $\boldsymbol\theta$ be the collection of model parameters
$\mathbf{m}, \boldsymbol\nu, \mathbf{a}, \mathbf{b}, \boldsymbol\lambda, \boldsymbol \rho$.
The complete log-likelihood for the model is
\begin{eqnarray}
  \ell (\mathbf{y}, \mathbf{u} ; \boldsymbol\theta) 
  & = &
  \sum_i \ell(\mathbf{y}_i, \mathbf{u}_i; \boldsymbol \theta) \\
  & = &
  \sum_i
  \log f (\mathbf{y}_i \mid \mathbf{u}_i ; \boldsymbol\theta) + \log f(\mathbf{u}_i; \boldsymbol\theta) 
  \\
  & = &
     -\frac{1}{2} \sum_{i,j} \log | 2 \pi R |
     -\frac{1}{2} \sum_{i,j}
     \left( \mathbf{y}_{ij} - Z_{ij} \mathbf{u}_i - \mathbf{b} \right)^T
     R^{-1}
     \left( \mathbf{y}_{ij} - Z_{ij} \mathbf{u}_i - \mathbf{b} \right) \nonumber
  \\
  & \phantom{=} &
    -\frac{1}{2} \sum_i \log | 2 \pi V |
    -\frac{1}{2} \sum_i
    \left( \mathbf{u}_{i} - \mathbf{m} \right)^T
     V^{-1}
     \left( \mathbf{u}_{i} - \mathbf{m} \right),
\end{eqnarray}
where
$Z_{ij}
  = 
  \mathbf{a} \mathbf{q}_{ij}^T$.
$\mathbf{y}_i$ is a vector of all biomarker measures stacked across all visits of individual $i$,
and $\mathbf{y}_{ij}$ is a vector of all biomarker measures for individual $i$ at visit $j$.
Given $n$ individuals, we use the 
summation notation $\sum_{i,j}$ as a shorthand for $\sum_{i=1}^n 
\sum_{j=1}^{v_i}$, where $v_i$ is the number of visits for individual $i$.

Since the marginal likelihood $f(\mathbf{y};\theta)$ involves an integral 
over all possible values of $\mathbf{u}_i$, maximizing it directly is difficult.
Therefore, we use the expectation-maximization~(EM) approach, where we consider 
the subject-specific variables $\{\mathbf{u}_i\}$ as hidden.
EM allows us to reformulate the maximization problem
in terms of the complete log-likelihood $\ell(\mathbf{y}, \mathbf{u}; \theta)$.
The observations $\mathbf{y}$
include biomarker measurements $\{\mathbf{y}_{ij}\}$ at each visit.
The unknown parameters are 
the subject-specific variable distribution parameters $\mathbf{m}$ and 
$\boldsymbol\nu$,
the trajectory parameters $\mathbf{a}$ and $\mathbf{b}$, 
and the noise covariance parameters $\boldsymbol \lambda$ and $\boldsymbol \rho$.

\subsection{E-step}
$(\mathbf{y}, \mathbf{u})$ are the complete data.
Let $\boldsymbol\theta' = \{\mathbf{m}', \boldsymbol\nu', \mathbf{a}', \mathbf{b}', \boldsymbol\lambda', \boldsymbol 
\rho'\}$ be the previous parameter estimates.
By Proposition~\ref{lemma:uhat2}, the E-step integral
is proportional to
$ 
\sum_i
  \int
  \Phi(\tilde{\mathbf{u}}_i; \hat{\mathbf{u}}'_i, \Sigma'_i)
  \ell \left( \mathbf{y}_i, \tilde{\mathbf{u}}_i ; \boldsymbol\theta\right)
  d\tilde{\mathbf{u}}_i
$, where
$\Phi$ is the multivariate normal probability density function with mean
\begin{equation}
\hat{\mathbf{u}}'_i = 
\left( \sum_j  Z'^T_{ij} R'^{-1} Z'_{ij} + V'^{-1} \right)^{-1}
\left( \sum_{j} Z'^T_{ij} R'^{-1}(\mathbf{y}_{ij} - \mathbf{b}') + V'^{-1} \mathbf{m}' 
\right),
\label{eq:uhat2}
\end{equation}
and covariance
$\Sigma'_i = \left( \sum_j  Z'^T_{ij} R'^{-1} Z'_{ij} + V'^{-1} \right)^{-1}$.
Note that for individuals with a single visit, $\sum_j  Z'^T_{ij} R'^{-1} Z'_{ij}$ 
is a singular matrix.
Considering $\mathbf{u}_i$ as parameters, as in~\citet{Jedynak2012} or 
\citet{Bilgel2015a}, is equivalent to assuming that $V$ has infinitely large 
diagonal elements (i.e., an uninformative uniform prior) such that its inverse disappears.
Therefore, it is not possible to compute 
$\hat{\mathbf{u}}'_i$ for individuals having only one visit using this 
approach.
On the other hand, incorporation of a bivariate normal prior on the subject-specific variables
$\mathbf{u}_i$ allows Eq.~\ref{eq:uhat2} to be computed for individuals with a 
single visit.

Evaluation of the E-step integral
involves second moments of a Gaussian 
random variable.
We ignore the terms that do not depend on
$\boldsymbol\theta$ as they will not be relevant in the maximization step
and obtain:
\begin{eqnarray}
  Q \left( \boldsymbol{\theta}, \boldsymbol{\theta}' \right)
  & = &
  -\frac{1}{2}\sum_{i,j} \log |R| \nonumber \\
  & \phantom{=} &
  -\frac{1}{2}\sum_{i,j}
  (\mathbf{y}_{ij} - Z_{ij}\hat{\mathbf{u}}'_i - \mathbf{b})^T R^{-1} (\mathbf{y}_{ij} - Z_{ij}\hat{\mathbf{u}}'_i- \mathbf{b})
  \nonumber \\
  & \phantom{=} &
  -\frac{1}{2} \sum_{i,j} \text{Tr} \left(  Z^T_{ij} R^{-1} Z_{ij} \Sigma'_i \right)
  -\frac{1}{2} \sum_{i} \log |V| \nonumber \\
  & \phantom{=} &
  -\frac{1}{2}\sum_{i}
  (\hat{\mathbf{u}}'_i - \mathbf{m})^T V^{-1} (\hat{\mathbf{u}}'_i - \mathbf{m}) 
  -\frac{1}{2} \sum_i \text{Tr} \left( V^{-1} \Sigma'_i \right) .
  \label{eq:Q}
\end{eqnarray}
   
\subsection{M-step}
\label{sec:EM-update-eq}
Here, we provide the update equations for the EM
algorithm obtained by maximizing $Q(\boldsymbol\theta, \boldsymbol\theta')$
with respect to each parameter, 
and provide derivations for these 
update equations in the Appendix.
The update equations depend on previous parameter estimates
$\boldsymbol\theta' = \{\mathbf{m}', \boldsymbol\nu', \mathbf{a}', \mathbf{b}', \boldsymbol\lambda', \boldsymbol 
\rho'\}$ as well as the progression score estimates
$s'_{ij} = \mathbf{q}^T_{ij} \hat{\mathbf{u}}'_i$, where $\hat{\mathbf{u}}'_i$ 
is as given in Eq.~\ref{eq:uhat2}:
      \begin{eqnarray}
  \mathbf{a} &=& 
  \frac{ \left(\sum_i v_i \right) \left( \sum_{i,j} \mathbf{y}_{ij} s'_{ij} \right) -
  \left( \sum_{i,j} \mathbf{y}_{ij} \right) \left( \sum_{i,j} s'_{ij} \right) }
  { \left(\sum_i v_i \right)
  \left( \sum_{i,j} \mathbf{q}^T_{ij} \Sigma'_i \mathbf{q}_{ij} +
  s'^2_{ij} \right) - \left( \sum_{i,j} s'_{ij} \right)^2
  } ,
\\
  \mathbf{b}
  &=&
  \frac{
  \left(
  \sum_{i,j}\mathbf{y}_{ij}
  \right)
  \left(
  \sum_{i,j} \mathbf{q}^T_{ij} \Sigma'_i \mathbf{q}_{ij} +
  s'^2_{ij}
  \right)
  -
  \left(
  \sum_{i,j} \mathbf{y}_{ij} s'_{ij}
  \right)
  \left(
  \sum_{i,j} s'_{ij}
  \right)
  }
  {
  \left(
  \sum_i v_i
  \right)
  \left(
  \sum_{i,j} \mathbf{q}^T_{ij} \Sigma'_i \mathbf{q}_{ij} +
  s'^2_{ij}
  \right)
  -
  \left(
  \sum_{i,j} s'_{ij}
  \right)^2
  } ,
  \\
  \mathbf{m} &=& \frac{1}{n} \sum_i \hat{\mathbf{u}}'_i ,
  \\
  \boldsymbol\nu &=& \argmax_{\boldsymbol \nu} Q(\boldsymbol \theta, \boldsymbol \theta') 
  ,
  \\
  \boldsymbol \lambda, \, \boldsymbol \rho 
  &=& \argmax_{\boldsymbol \lambda, \boldsymbol \rho} Q(\boldsymbol \theta, \boldsymbol \theta') 
  .
\end{eqnarray}
Note that if $C$ is fixed to be the identity matrix,
a closed form solution for
$\boldsymbol \lambda$ exists, as given in
Equation~\ref{eq:multiLambda}.
Once the optimal parameters are found, the subject-specific variables are predicted
using Eq.~(\ref{eq:uhat2}).

\subsection{Parameter standardization}
As described in Proposition~\ref{lemma:param-standardization}, there are certain
reparameterizations that yield identical models. 
For example, one can multiply the trajectory slope parameters by 2 and divide all progression
scores by 2 (which is achieved by dividing all $\alpha$ and $\beta$ values by 2) without altering
the model.
This is the scaling degree of freedom. 
There is also a translation degree of freedom. 
We account for these degrees of freedom and anchor the model by calibrating the progression 
score scale. 
We calibrate such that baseline PS has a mean of 0 and a standard deviation of 1. 
This involves replacing the model parameters 
$\left\{ \mathbf{m}, V(\boldsymbol\nu), \mathbf{a}, \mathbf{b},
R(\boldsymbol\lambda, \boldsymbol \rho) \right\}$
with
$\left\{w\mathbf{m}+
\begin{bmatrix}
  0 \\ z
\end{bmatrix}, w^2 V(\boldsymbol\nu), \frac{1}{w} \mathbf{a}, \mathbf{b} - \frac{z}{w} \mathbf{a}, 
R (\boldsymbol\lambda, \boldsymbol \rho)
\right\}$,
where
$-\frac{z}{w} = \frac{1}{n} \sum_i s_{i1}$ is the mean progression score at 
baseline,
and $\frac{1}{w} = \sqrt{\frac{1}{n} \sum_i \left(  s_{i1} - \frac{1}{n} \sum_i s_{i1} \right)^2}$ 
is the standard deviation of baseline progression scores.
We standardize the subject-specific estimates $\alpha_i, \beta_i$, and $s_{ij}$ 
accordingly:
$\alpha^*_i = w\alpha_i, \beta^*_i = w \beta_i + z$, and $s^*_{ij} = w s_{ij} + 
z$.
This reparametrization yields a PS scale where 0 corresponds to the sample average at 
baseline and the variance of PS at baseline is 1.

\subsection{Implementation details}
We first fit the model assuming that $C=I_{K \times K}$ and $\Lambda$ is a diagonal matrix
with positive elements $\boldsymbol \lambda$.
We denote the estimated $\Lambda$ in this model as $\hat{\Lambda}$.
The estimates obtained from this model for the parameters
$\mathbf{a}, \mathbf{b}, \mathbf{m}, V$
are used as initializations in the model where $C=C(\rho)$.
In this model where correlations are taken into account, 
we assume that $\Lambda = \lambda \hat{\Lambda}$, where $\lambda$ 
is an unknown parameter to be estimated.
The spatial correlation function among those presented in Table~\ref{tab:corr-fun} 
that results in the highest log-likelihood value is chosen for the final model.

The EM algorithm is implemented in MATLAB~8.1
and Statistics Toolbox~8.2 (The MathWorks Inc., Natick, MA).
Our code is freely available
online.\footnote{\url{https://www.iacl.ece.jhu.edu/Resources}}

\subsection{Confidence intervals}
We use bootstrapping via Monte Carlo resampling to estimate confidence intervals 
for each model parameter.
We sample with replacement from the original collection of subjects to generate 
a new dataset containing an equal number of subjects and fit the model on this generated 
sample.
This sampling and fitting procedure is repeated to generate bootstrap estimates.
We then compute 95\% confidence intervals for each parameter across the 
bootstrap estimates.
In the bootstrap experiments, we fix the value of $\rho$ at its estimate on the 
whole sample to enable faster computation.

\subsection{Comparison to linear mixed effects model}
We compared our model to a linear mixed effects~(LME) model that included 
random intercepts and slopes at each voxel.
The LME model 
for subject~$i$, visit~$j$ and voxel~$k$
is given by
\begin{equation}
  y_{ijk} = (\eta_k + \eta_{ik}) t_{ij} + (\gamma_k + \gamma_{ik}) + \varepsilon_{ijk} , 
\end{equation}
where
$\begin{pmatrix}
\eta_{k} \\
\gamma_{k}
\end{pmatrix}$ are the fixed effects,
$\begin{pmatrix}
\eta_{ik} \\
\gamma_{ik}
\end{pmatrix} \sim \mathcal{N}(0,\Xi_k)$ are the random effects and
$\varepsilon_{ijk} \sim \mathcal{N}(0,\sigma^2_k)$ is the observation noise.
We used the LME implementation in MATLAB Statistics Toolbox 8.2.

\subsection{Simulated data set}
We simulated visits such that the sample was similar to our PET 
data in terms of number of visits per subjects and age range.
We generated a data set with 100 individuals, each with up to 7 visits
with $5 \times 5 \times 5$ images with 4 mm isotropic voxels.
We fixed the ground truth values of the model parameters
$\boldsymbol\theta = \{
\mathbf{m}, \boldsymbol\nu, \mathbf{a}, \mathbf{b}, \boldsymbol\lambda, \rho\}$
at values close to those we observed in exploratory models fitted to DVR data.
We generated $\mathbf{u}_i$ from a bivariate normal distribution with mean
$\mathbf{m}$ and variance $V(\boldsymbol\nu)$.
The progression score for each visit was then computed as $s_{ij} = \mathbf{q}^T_{ij} 
\mathbf{u}$, and observations were generated using the PS model.
We performed 1000 bootstrap iterations to obtain 95\% confidence intervals for 
each parameter and subject-specific variable.
We computed the cosine similarity for each variable for each bootstrap experiment
using
\begin{equation}
  \frac{\boldsymbol\phi^T \widehat{\boldsymbol\phi}}
  { \| \boldsymbol\phi \|_2 \| \widehat{\boldsymbol \phi} \|_2 },
\end{equation}
where $\widehat{\boldsymbol\phi}$ is the estimate of the variable of interest 
($\mathbf{a}, \mathbf{b}, \boldsymbol\alpha, \boldsymbol\beta,$ or $\mathbf{s}$) 
and $\boldsymbol\phi$ is the corresponding ground truth value.
A value of 1 indicates a perfect estimate.

\subsection{Amyloid imaging data set}
We used longitudinal positron emission tomography~(PET) data for participants 
from the Baltimore Longitudinal Study of Aging~\citep{Shock1984} neuroimaging 
substudy~\citep{Resnick2000}.
Participant demographics are presented in Table~\ref{tab:demographics}.
Starting in 2005,
amyloid-beta~(A$\beta$) PET scans were acquired on a GE Advance scanner
over 70 minutes following an intravenous 
bolus injection of Pittsburgh 
compound~B~(PiB), yielding dynamic PET scans with 33 time 
frames.
PET images were reconstructed using filtered backprojection with a ramp filter,
yielding a spatial resolution of approximately 4.5 mm full width at half max at 
the center of the field of view (image matrix = $128 \times 128$, 35 slices,
pixel size = $2 \times 2$ mm, slice thickness = 4.25 mm).

The frames of each dynamic PiB-PET scan were aligned to the average of the first two minutes
to remove motion~\citep{Jenkinson2002}.
For registration purposes, we obtained static images by averaging the first 20 minutes
of the dynamic PiB-PET scan.
Follow-up PiB-PET scans were rigidly registered onto the baseline PiB-PET within each participant using 
the 20-minute average images.
Baseline magnetic resonance images (MRIs) were rigidly registered onto 
their corresponding 20-minute PiB-PET average, and their FreeSurfer
segmentations~\citep{Dale1999,Desikan2006} were transformed accordingly.
Distribution volume ratio~(DVR) images were calculated in the native space of 
each PiB-PET image using the simplified reference tissue
model with the cerebellar gray matter as reference tissue~\citep{Zhou2003}.
The MRIs coregistered with the PET were deformably registered~\citep{Avants2008} onto a
study-specific template~\citep{Avants2010,Bilgel2015} and transformed
to 4 mm isotropic MNI space using a pre-calculated affine transformation.
The resulting mappings were applied to the DVR
images that have been
registered to baseline to bring them into the MNI space.
We used all voxels within the brain mask in the MNI space to fit the PS model.

Mean cortical DVR is a widely used measure obtained from PiB-PET images for
quantifying the level of brain amyloid.
We computed mean cortical DVR for each PiB-PET 
image by averaging the voxelwise DVR values across cingulate, frontal, parietal, lateral 
temporal, and lateral occipital cortices, excluding the sensorimotor strip.
We used a mean 
cortical DVR threshold of 1.06,
which was computed from a two-class Gaussian mixture model
fitted on baseline measures, to separate individuals into PiB- and PiB+ 
groups, as described in~\citet{Bilgel2015b}.

\begin{table}
  \centering
  \caption{Participant demographics. MCI = mild cognitive impairment, 
  SD = standard deviation.}
  \begin{tabular}{l|l}
    Characteristic & $N = 104$ \\
    \toprule
    Baseline age in years, mean (SD) & 77.0 (7.9) \\
    Range & 55.7--93.4 \\
    \hline
    Female, n (\%) & 48 (46\%) \\
    \hline
    PiB-PET scans, n & 300 \\
    \hline
    PiB-PET per subject, n & 2.9 (1.8) \\
    Range & 1--7 \\
    \hline
    Years between first and last scan, mean (SD) & 3.3 (2.9) \\
    Range & 0.0--9.0 \\
    \hline
    Mean cortical DVR at last visit (SD) & 1.12 (0.19) \\
    \hline
    PiB+ at last visit, n (\%) & 42 (40\%) \\
    \hline
    MCI diagnosis at last visit, n (\%) & 3 (3\%) \\
    \hline
    Dementia diagnosis at last visit, n (\%) & 4 (4\%) \\
    \bottomrule
  \end{tabular}
  \label{tab:demographics}
\end{table}

\subsection{Hypothesis testing}
The confidence intervals obtained via bootstrapping allow for hypothesis testing.
For the amyloid images, 
we focus on studying the precuneus, since previous cross-sectional
amyloid imaging studies have suggested
that precuneus has the highest deposition levels~\citep{Mintun2006} and provided
preliminary evidence that it may be the most rapid accumulator~\citep{Rodrigue2012}
among cortical regions.
Our specific hypotheses are as follows:
\begin{enumerate}
  \item The precuneus has the highest amyloid load along stages of amyloid accumulation.
  \item The precuneus accumulates amyloid faster than other cortical regions.
\end{enumerate}
We refer to the progression scores calculated using the DVR images
as A$\beta$-PS.
To test the first hypothesis, we compare the amyloid levels in the precuneus with other regions
at various A$\beta$-PS values spanning the range observed in our data set.
Our purpose in performing these comparisons is to shed light onto the temporal 
ordering of changes in different cortical regions.
For each bootstrap experiment, we average the predicted DVR levels
$\hat{y}_k (s) = a_k s + b_k$ within each ROI to obtain
\begin{equation}
  \hat{y}_r(s) = \frac{1}{\left| \text{ROI}_r \right|} \sum_{k \in \text{ROI}_r} 
  \hat{y}_k(s),
\end{equation}
where $k$ is the voxel index and $r$ is the ROI index.
We then compute the following statistic for each bootstrap:
\begin{equation}
  T = \hat{y}_\text{precuneus} (s) - \max_{r \neq \text{precuneus}} \hat{y}_r (s) .
\end{equation}
We reject the null hypothesis that the amyloid load in the precuneus at 
A$\beta$-PS~$= s$
is not different from other
regions at significance level $\gamma$ if the two-sided
$100(1-\gamma)$ confidence interval of the test statistic $T$, computed
using the bootstrap estimates of $T$, does not contain $0$.
The smallest value of $\gamma$ such that the two-sided $100(1-\gamma)$ confidence
interval of the test statistic $T$ contains $0$ is the $p$-value of the test.

To test the second hypothesis, we pursue a similar approach using the trajectory slope parameter
$\mathbf{a}$.
For each bootstrap experiment, we obtain the average $a_k$ per cortical ROI to obtain
$a_r$, where $r$ the ROI index.
We then compute the following statistic for each 
bootstrap:
\begin{equation}
  T = a_\text{precuneus} - \max_{r \neq \text{precuneus}} a_r .
\end{equation} 
The condition for the rejection of the null hypothesis
that the rate of amyloid accumulation in the precuneus is not 
different from that of other regions is as described previously.

\section{Results}

\subsection{Simulation}
The Akaike information criterion (AIC) was $39.0 \times 10^3$ for the LME model,
$32.0 \times 10^3$ for the PS model where $C=I_{K \times K}$, and 
$-1.6 \times 10^3$ for the PS model where $C=C(\rho)$, 
indicating that the PS model where spatial correlations are modeled fits the data the best.
We computed the percentage of variables that are ``correct'' by counting the
variables whose ground truth values fell within the 95\% confidence interval computed via 
bootstrapping.
For example, for $\mathbf{a}$ and $\mathbf{b}$, a value of 10\% indicates that 
10\% of the biomarkers had a correct estimate for $a_k$ and $b_k$.
For $\boldsymbol\alpha$ and $\boldsymbol\beta$, 10\% 
indicates that 10\% of the individuals had a correct estimate for $\alpha_i$ and 
$\beta_i$.
For $\mathbf{s}$, 10\% indicates that 10\% of the visits had a 
correct estimate for $s_{ij}$.
Simulation results are presented in Table~\ref{tab:simulation}.
When there are correlations in the data, not modeling them yields inaccurate 
estimates for the trajectory slope parameter $\mathbf{a}$,
the subject-specific variables $\alpha_i, \beta_i$, and the 
progression scores $s_{ij}$.
Using the correlation model improves these estimates significantly.

\begin{table}[h!]
\centering
\caption{Simulation results. Mean cosine similarity values across 1000 bootstrap experiments 
(with standard deviation) and percentage of variable elements that are correct based on
95\% confidence intervals are presented.}
\setlength{\tabcolsep}{10pt}
\begin{tabular}{r|cc|cc}
  & \multicolumn{2}{c}{$C = I_{K \times K}$} & \multicolumn{2}{c}{$C = C(\rho)$} \\
  \hline
  Variable & Cosine similarity & \% correct & Cosine similarity & \% correct \\
  \hline
  $\mathbf{a}$ & $0.9514 \pm 0.0154$ & 24 & $0.9821 \pm 0.0073$ & 98 \\
  $\mathbf{b}$ & $0.9998 \pm 0.0001$ & 99 & $0.9998 \pm 0.0001$ & 98 \\
  $\boldsymbol\alpha$ & $0.1587 \pm 0.0812$ & 18 & $0.7922 \pm 0.0332$ & 40 \\
  $\boldsymbol\beta$ & $0.1250 \pm 0.0794$ & 19 & $0.7835 \pm 0.0339$ & 38 \\
  $\mathbf{s}$ & $0.8017 \pm 0.0443$ & 23 & $0.9881 \pm 0.0053$ & 85 \\
  \bottomrule
\end{tabular}
\label{tab:simulation}
\end{table}

\begin{figure}
  \centering
  \includegraphics[width=.9\textwidth]{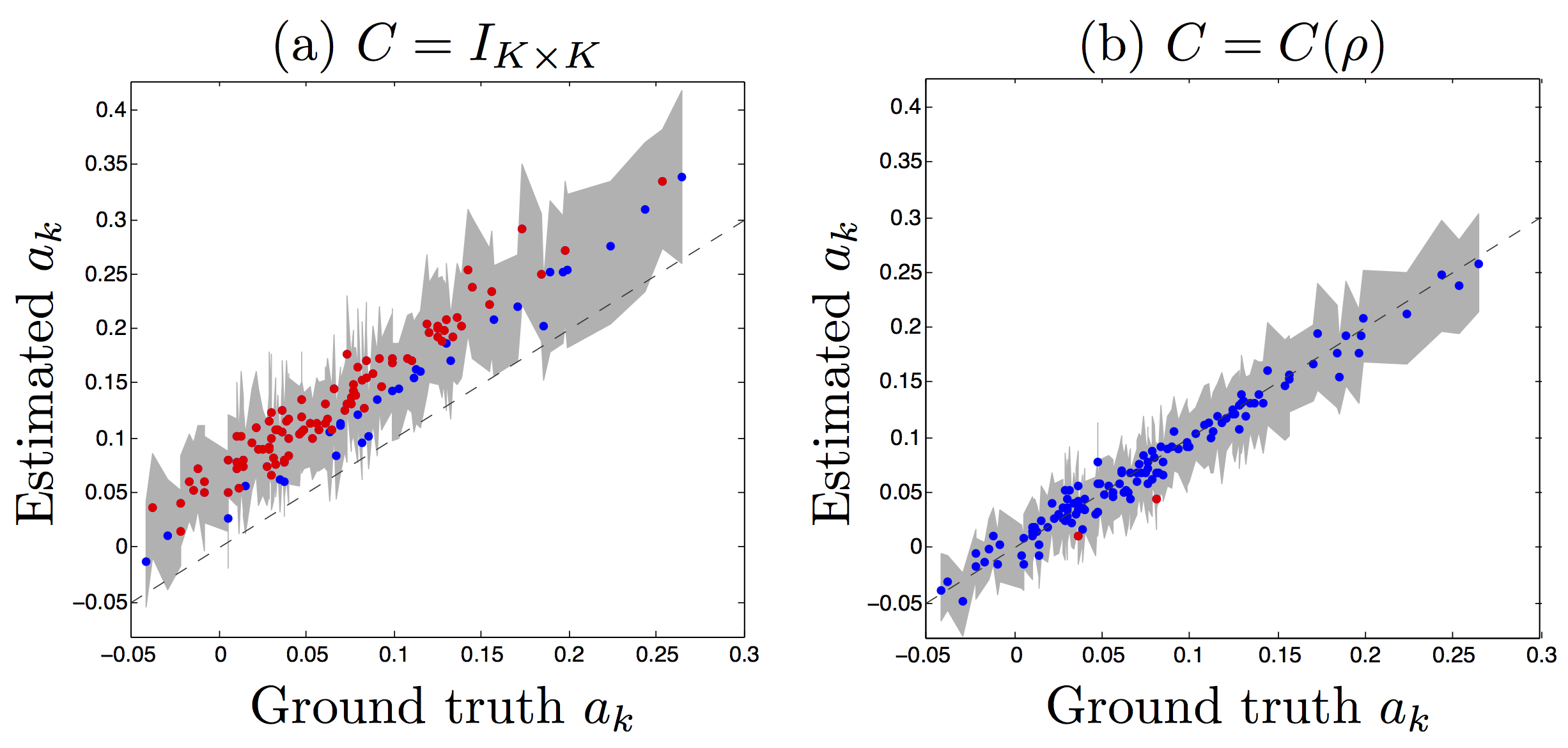}
  \caption{Estimated trajectory slope parameters $a_k$ vs ground truth. Dashed line indicates
  $x=y$. Gray band corresponds to the 95\% confidence intervals obtained from bootstrapping.
  Estimates are shown in blue if their 95\% confidence interval intersects the $x=y$ line, 
  and in red otherwise. 
  Results from the model where (a) $C = I_{K \times K}$ and (b) $C = C(\rho)$.}
\end{figure}
\begin{figure}
  \centering  
  \includegraphics[width=.9\textwidth]{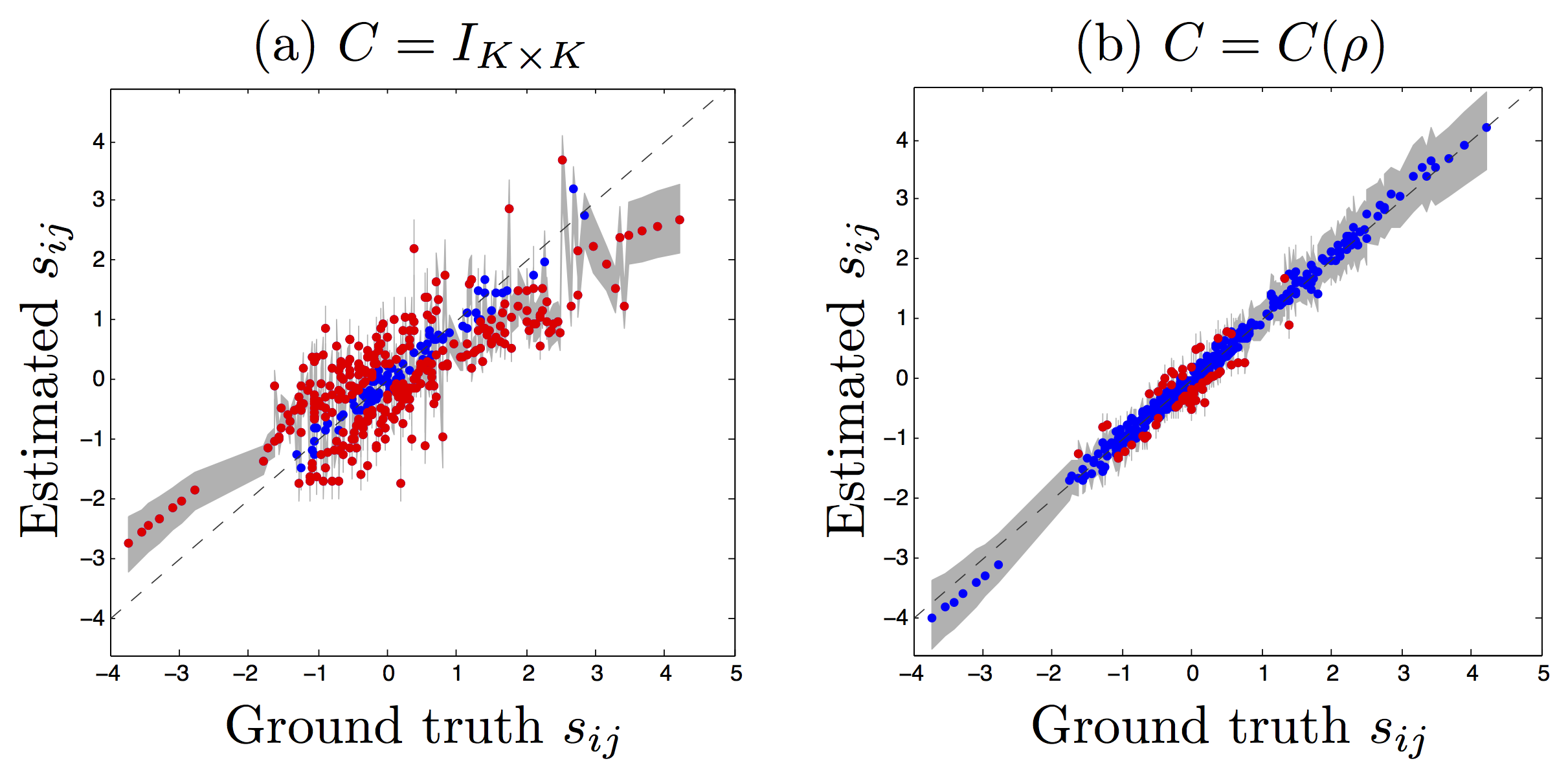}
  \caption{Predicted progression scores $s_{ij}$ vs ground truth. Dashed line indicates
  $x=y$. Gray band corresponds to the 95\% confidence intervals obtained from bootstrapping.
  Estimates are shown in blue if their 95\% confidence interval intersects the $x=y$ line, 
  and in red otherwise. 
  Results from the model where (a) $C = I_{K \times K}$ and (b) $C = C(\rho)$.}
\end{figure}

\subsection{Amyloid images}
In our preliminary analysis of the noise spatial correlation structure using the 
semivariogram~\citep{Cressie1980,Bilgel2015a},
we observed that the rational quadratic had the best fit to 
the empirical semivariogram (Inline Supplementary Fig.~\ref{fig:semivariogram}).
Rational quadratic also yielded the 
fit with largest log-likelihood, and thus was selected as the correlation function 
for the model.
AIC was $-1.2 \times 10^7$ for the LME model,
$-1.1 \times 10^7$ for
the model where $C=I_{K \times K}$, and $-2.0 \times 10^7$ for the model
where $C = C(\rho)$, indicating that the PS model where spatial correlations are modeled fits
the data the best.
The spatial correlation parameter $\rho$ for the rational quadratic
correlation model was estimated to be 4.5 mm, and $\lambda$ was estimated to be 0.929.
A comparison of the PS values obtained from the different 
correlation models are presented in the
Supplementary Material (Inline Supplementary Fig.~\ref{fig:bland-altman-ps}).
Model fitting using 104 subjects with a total of 300 visits
each having about $30,000$ brain voxels
took approximately 5 seconds 
on an Intel Xeon 8-core 3.1 GHz machine with 128 GB RAM
for the case where voxels are assumed to be independent (i.e., $C=I_{K \times K}$).
The following model fitting with $C=C(\rho)$ took approximately 30 minutes per 
EM iteration, with convergence achieved in 2 to 6 iterations depending on the 
correlation structure.

\subsubsection{Comparison of A$\beta$-PS to mean cortical DVR}
\begin{figure}
  \centering
  \includegraphics[width=.9\textwidth]{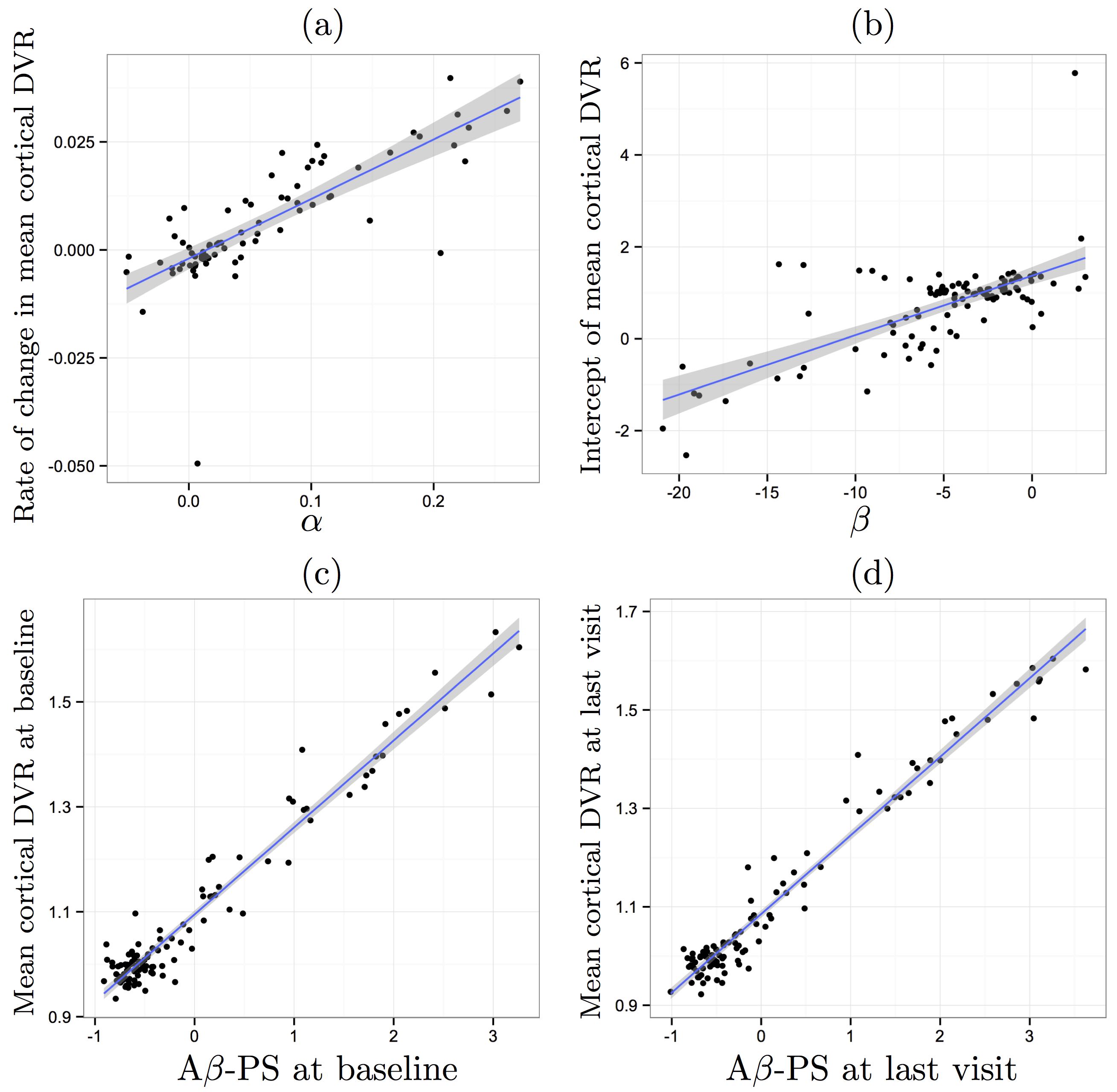}
  \caption{Correlation of estimated subject-specific variables with 
  mean cortical DVR measures.
  The line of best fit is shown in blue, and its 95\% confidence band in 
  gray.
  (a)~Rate of annual change in mean cortical DVR vs. 
  $\alpha$, the predicted rate of change in amyloid progression score ($R^2 = 0.62$).
  (b)~Intercept of mean cortical DVR vs. $\beta$,
  the progression score intercept ($R^2 = 0.48$).
  (c)~Mean cortical DVR vs. A$\beta$-PS at baseline ($R^2 = 0.95$).
  (d)~Mean cortical DVR vs. A$\beta$-PS at last visit ($R^2 = 0.96$).}
  \label{fig:ps-vs-mcdvr}
\end{figure}

We compared the subject-specific variables obtained from model fitting to 
empirical values obtained from mean cortical DVR, which is the widely used 
measure for quantifying levels of brain amyloid~(Fig.~\ref{fig:ps-vs-mcdvr}).
For each individual with at least two visits, we fit a line to the mean 
cortical DVR data to estimate the slope of amyloid accumulation as well as the intercept.
The subject-specific variable $\alpha_i$, which represents the rate of amyloid 
progression, explained $62\%$ of the variability in the empirical slope of amyloid 
accumulation computed from longitudinal mean cortical DVR.
We observed a much higher correlation between mean cortical DVR and A$\beta$-PS
($R^2 = 0.95$ at baseline and $0.96$ at last visit).

\begin{figure}
  \centering
  \includegraphics[width=.9\textwidth]{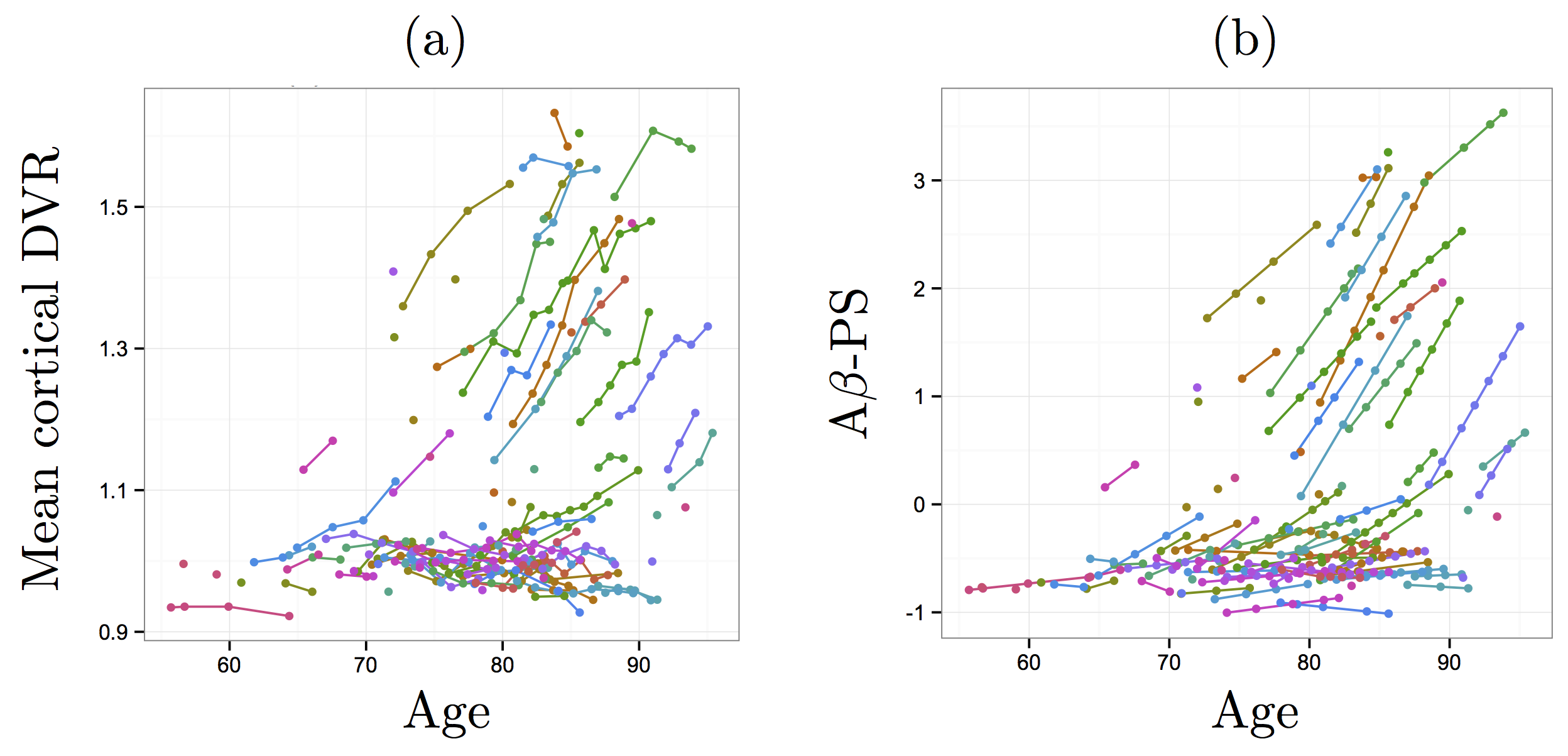}
  \caption{(a) Mean cortical DVR and (b) A$\beta$-PS plotted against age. 
  Longitudinal data points are connected by lines within each subject.
  Different colors indicate different subjects.
  }
  \label{fig:ps-vs-age}
\end{figure}

When plotted against age, mean cortical DVR and A$\beta$-PS revealed similar 
patterns~(Fig.~\ref{fig:ps-vs-age}).
The progression as measured using voxelwise DVR data is not linearly associated 
with age, and the A$\beta$-PS was able to capture this.

\subsubsection{Amyloid trajectories}
\begin{figure}
  \includegraphics[width=1\textwidth]{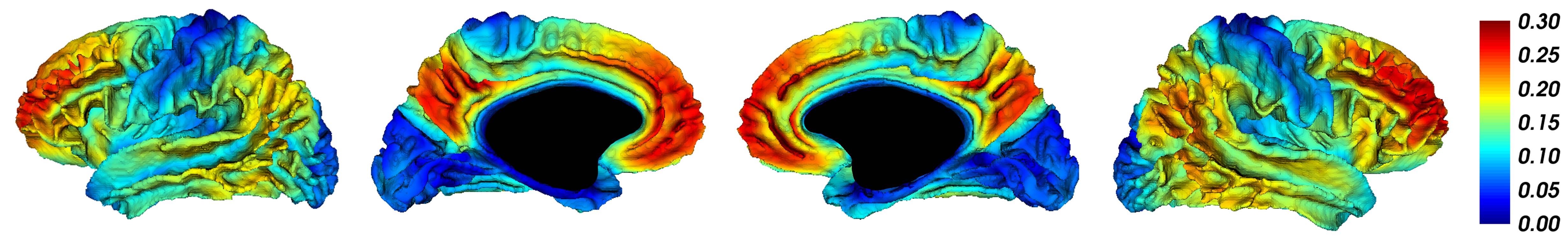}
  \caption{Slope parameters 
  $a_k$ obtained from 
  voxelwise
  PS model projected
  onto the cortical surface.
  For each unit increase in A$\beta$-PS, the DVR value at voxel $k$
  increases by $a_k$.
  }
  \label{fig:lme-ps-A}
\end{figure}

\begin{figure}
  \centering
  \includegraphics[width=1\textwidth]{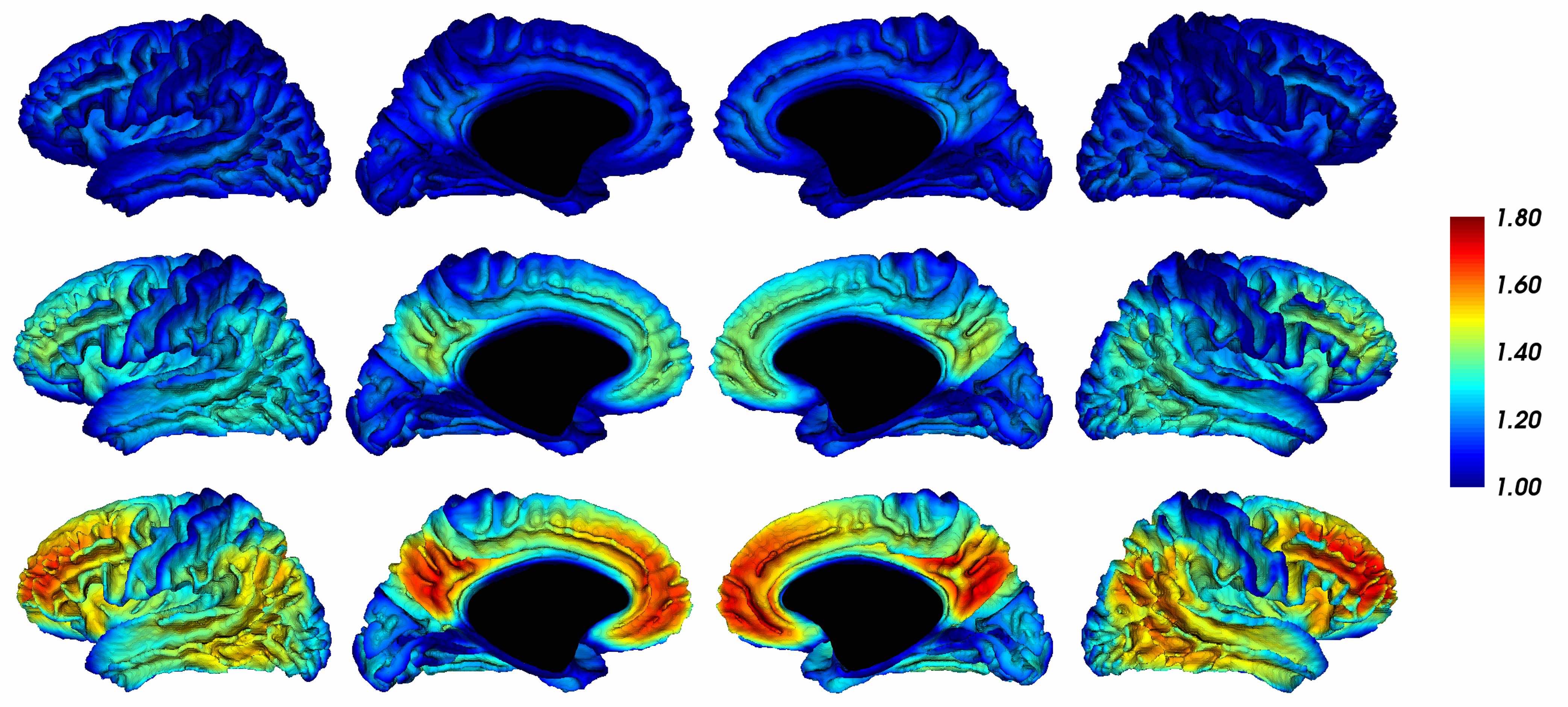}
  \caption{Predicted DVR levels at
  A$\beta$-PS $=-0.6$ (\emph{top}), 0.4 (\emph{middle row}), and 1.5 (\emph{bottom}).}
  \label{fig:predDVR}
\end{figure}

The trajectory slope parameters $a_k$ obtained from the
PS model~(Fig.~\ref{fig:lme-ps-A})
revealed symmetric patterns across the cerebral hemispheres.
The precuneus and frontal lobe showed
the greatest increases in DVR with A$\beta$-PS, smaller increases in lateral 
temporal and temporoparietal regions, and minimal increases in the occipital 
lobe and the sensorimotor strip.
Voxelwise trajectories are further illustrated in Fig.~\ref{fig:predDVR},
which shows the predicted DVR values on the cortical surface at three A$\beta$-PS 
levels.

\begin{figure}
  \centering
  \includegraphics[width=.7\textwidth]{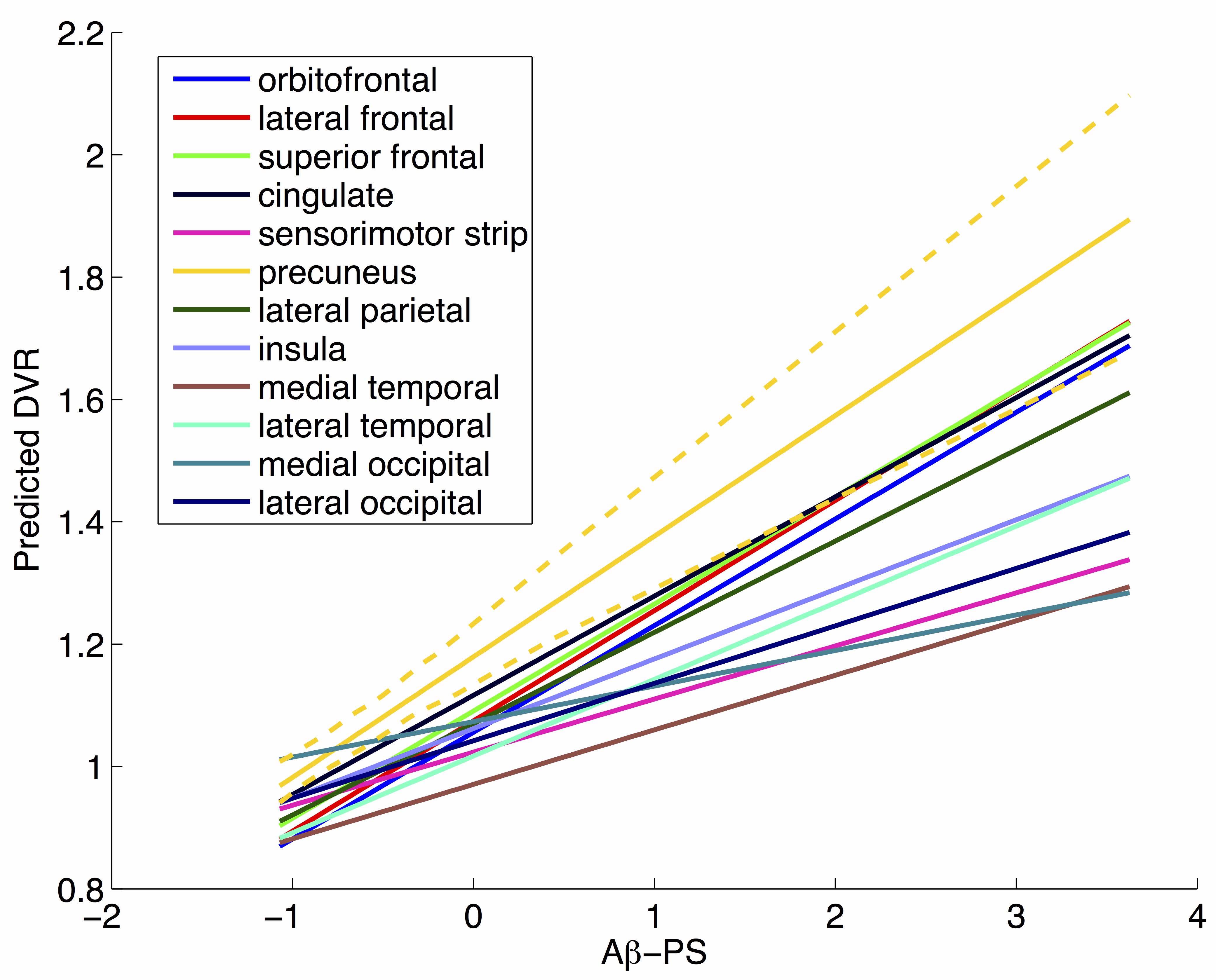}
  \caption{Regional trajectories as function of A$\beta$-PS.
  The PS model was used to make voxelwise predictions at a range of
  A$\beta$-PS values, and these predictions were averaged within each ROI to obtain regional
  trajectories.
  The dashed lines indicate the 95\% confidence band
  obtained using bootstrap results
  for the precuneus.}
  \label{fig:precuneus-trajectory-CI}
\end{figure}

\begin{figure}
  \centering
  \includegraphics[width=1\textwidth]{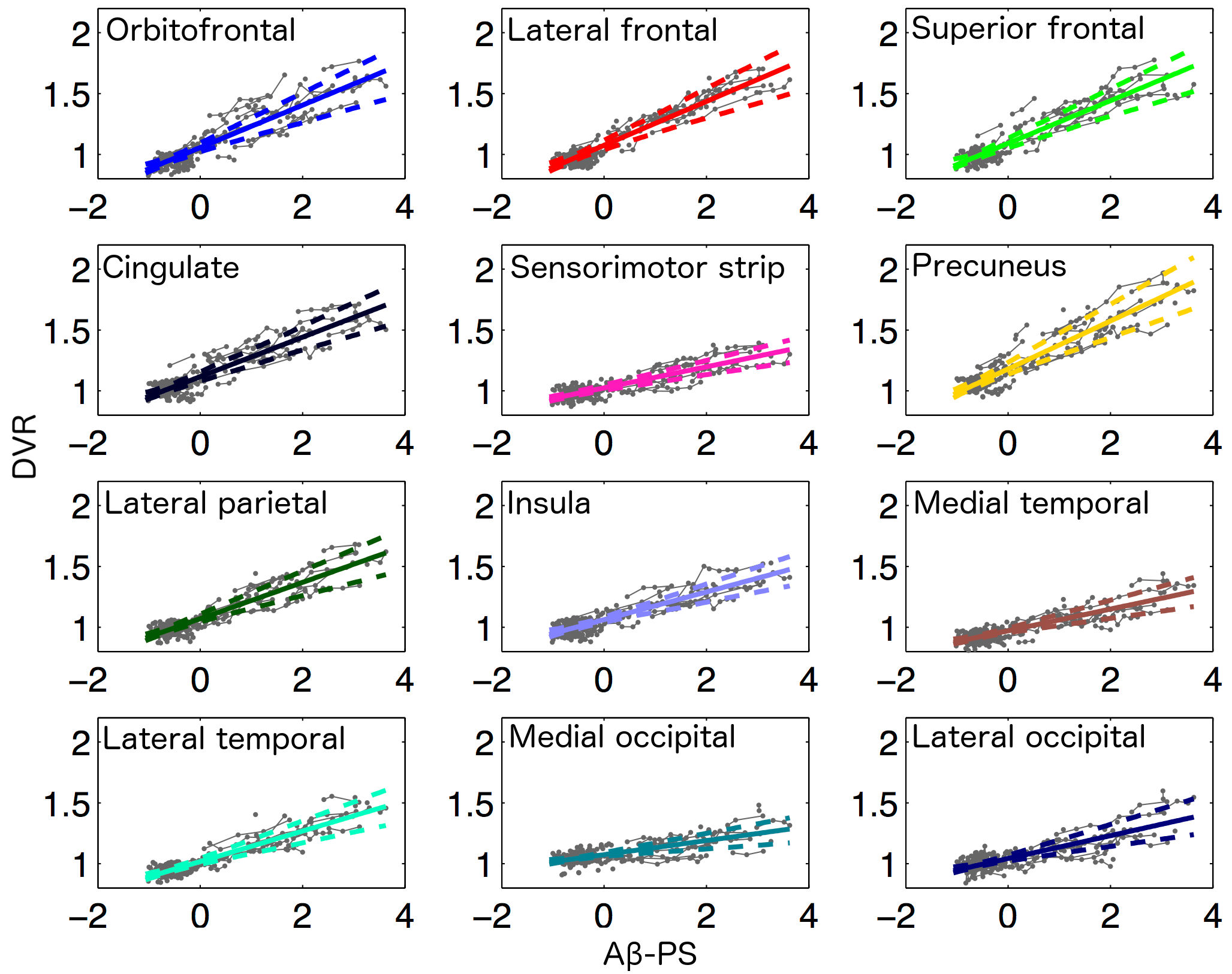}
  \caption{Regional trajectories as function of A$\beta$-PS.
  The PS model was used to make voxelwise predictions at a range of
  A$\beta$-PS values, and these predictions were averaged within each ROI to obtain regional
  trajectories.
  The dashed lines indicate the 95\% confidence bands for the cortical regions.
  Estimated trajectories with their 95\% confidence bands are superimposed on 
  observed longitudinal data (in gray).}
  \label{fig:ROI-trajectory-CI}
\end{figure}

In order to better investigate regional trends, 
we averaged the PS model results within each cortical ROI and plotted these ROI averages
as a function of A$\beta$-PS~(Fig.~\ref{fig:precuneus-trajectory-CI}).
We used bootstrap results to compute 95\% confidence bands for these
ROI averages.
Based on the 95\% confidence band of the precuneus illustrated in
Fig.~\ref{fig:precuneus-trajectory-CI}, precuneus appears to be the earliest 
accumulator and has the highest amyloid levels through late stages of amyloid
accumulation (A$\beta$-PS$\geq 2$).
We present confidence bands for other cortical regions in Fig.~\ref{fig:ROI-trajectory-CI}.

\begin{figure}
  \centering
  \includegraphics[width=1\textwidth]{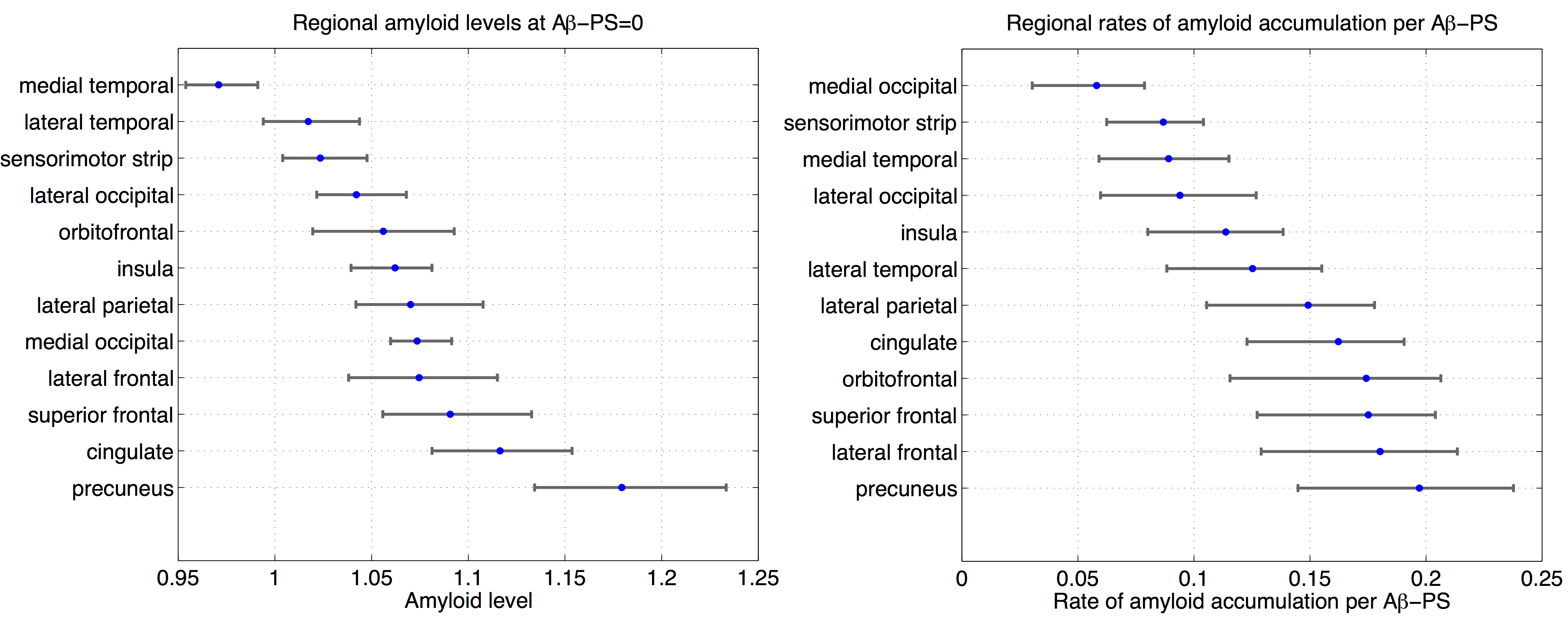}
  \caption{Comparison of levels of amyloid at A$\beta$-PS$=0$ and rates of amyloid accumulation
  across cortical regions.
  The intercept parameter $\mathbf{b}$ obtained from the PS model was
  averaged within each ROI to obtain the regional amyloid levels at A$\beta$-PS$=0$, and 
  the trajectory slope parameter $\mathbf{a}$ was averaged within each ROI to obtain regional
  rates.
  }
  \label{fig:comparison-ROI}
\end{figure}

Based on our hypothesis testing procedure, we found that precuneus 
had the highest amyloid levels at
A$\beta$-PS values of -0.5, 0, 1, 2, and 3 (all $p<0.01$).
On the other hand, while the estimated rate of amyloid accumulation was highest 
in the precuneus, this was not statistically significant ($p=0.33$).
A comparison of the levels of amyloid at A$\beta$-PS$=0$ and rates of 
accumulation across ROIs is presented in Fig.~\ref{fig:comparison-ROI}.

\subsubsection{Comparison to LME results}
The trajectories we obtained using the PS model were consistent with the results of the 
LME model (Inline Supplementary Figs.~\ref{fig:lme-A},~\ref{fig:precuneus-trajectory-CI-lme}).
However, the fixed effects obtained from the LME model did not describe the 
observed data as well as the trajectory slopes obtained from the PS model
(Inline Supplementary Fig.~\ref{fig:ROI-trajectory-CI-lme}).
Using the LME model estimates and a bootstrapping procedure similar to the one we applied for 
the PS model, we found that the precuneus had the highest regional amyloid levels 
compared to other cortical regions at ages 70, 80, and 90 (all $p<0.01$),
but not at age 65 ($p=0.12$).
Similar to our findings based on the PS model, 
the rate of amyloid accumulation was highest in the precuneus, but this was not statistically 
significant ($p=0.75$).
A comparison of the levels of amyloid at the mean baseline age of the sample and rates of 
accumulation across ROIs is presented in Inline Supplementary
Fig.~\ref{fig:comparison-ROI-lme}.

\section{Discussion}
We presented a statistical model for estimating longitudinal trajectories of voxelwise
neuroimaging data. Our model is based on the concept that age is not a good metric for
disease progression since each individual has his or her own onset and rate of progression 
of disease. 
We accounted for these inter-individual differences to temporally align individuals based on 
their collection of voxelwise measurements. 
As a result of this temporal alignment, we obtained a metric of disease or underlying biological 
process stage as reflected in the neuroimaging data, and we call this metric ``progression
score''.
By analyzing voxelwise neuroimaging data as a function of progression score rather than time, 
we constructed trajectories that better represent changes occurring with disease stage.

Simulation results showed that the progression score model fits the data
better than the linear mixed effects model, as evidenced by AIC. 
Furthermore, the results showed that
modeling spatial correlations is 
important for extracting accurate summary scores from data with underlying 
correlations.
In the absence of correlation modeling when the underlying data are correlated,
estimates of the trajectory slopes $\mathbf{a}$ as well as subject-specific variables 
$\alpha, \beta$, and $s$ were adversely affected.
The difference in the subject-specific variables 
was mainly due to inaccurate estimates of the prior 
covariance $V$.
Note that
\begin{equation}
  \text{Cov}(\mathbf{y}_{ij}) = \mathbf{a} \mathbf{q}^T_{ij} V \mathbf{q}_{ij} 
  \mathbf{a}^T + R,
\end{equation}
which explains why both $\mathbf{a}$ and $V$ are affected when correlations 
are not modeled through the covariance matrix $R$.
This highlights the importance of modeling spatial correlations in data where 
such effects are evident, especially in order to estimate correct trajectory 
slopes and individualized progression scores.
To further understand this phenomenon, we conducted simulations (data not presented)
where we duplicated biomarkers with lower signal-to-noise ratios and repeated 
model fitting including these duplicate biomarkers. We observed that the parameter estimates 
as well as the PS values were biased if we assumed independence across biomarkers. 
This might be because the model interprets each of the duplicated biomarkers as a separate 
piece of evidence towards the computation of individual progression scores. 
However, when we included a proper noise correlation model, we were able to recover our 
original results on the collection of unique biomarkers.

The proposed EM framework simplifies the estimation of 
spatial correlation parameters, and accurately estimates the trajectory 
parameters $\mathbf{a}$ and $\mathbf{b}$ as well as predicting the progression scores $s$.
The performance of the fitting procedure in predicting $\alpha$ and $\beta$
is worse than for $s$.
This may be due to the modeling of $\alpha$ and $\beta$ as
random variables that do not directly contribute to the observations $\mathbf{y}$.
We observed a similar pattern in the amyloid data: the agreement between
estimated A$\beta$-PS and mean cortical DVR was greater than the agreement
between estimated $\alpha$ and the rate of change of mean cortical DVR per 
individual.
This suggests that the progression scores are more reliable than the 
subject-specific variables.
A$\beta$-PS was highly correlated with mean cortical DVR, a widely used measure 
for quantifying PiB-PET scans and assessing longitudinal change. This indicates 
that A$\beta$-PS is a meaningful score extracted from voxelwise imaging data.
Unlike mean cortical DVR, there are no \emph{a priori} assumptions regarding which
regions or voxels should be included in the computation of A$\beta$-PS.
This property of PS allows for the discovery of new patterns in longitudinal data.

Trajectories obtained from the PS model captured a wider dynamic range of DVR
values and had a better fit to the observed data
compared to the trajectories obtained using the LME model fixed effects.
This difference between the PS and LME models underscores the fact that age is not an 
appropriate metric for staging amyloid accumulation. 
By aligning individuals based on their amyloid scans, the PS model enables a better temporal 
metric of amyloid staging.

A$\beta$-PS allows for the exploration of longitudinal 
voxelwise trajectories within a hypothesis testing framework due to its underlying
statistical model.
Our results suggest that the precuneus exhibits the earliest cortical amyloid
changes, but that its rate of amyloid
accumulation does not differ significantly from other cortical regions.
Based on a qualitative evaluation of our estimated trajectories, we found that amyloid 
accumulation in the precuneus is followed by cingulate and frontal cortices, 
then by lateral parietal cortex, followed by insula and lateral temporal cortex.
We observed minimal amyloid accumulation in visual cortex, hippocampal formation
and the sensorimotor strip, which agrees with previous reports
that these regions accumulate amyloid in later stages of AD~\citep{Braak1991}.
Previous reports have highlighted precuneus as an early amyloid 
accumulator~\citep{Mintun2006,Rodrigue2012} as well as frontal, cingulate, and 
parietal regions~\citep{Jack2008}.
Contrary to our finding 
highlighting precuneus as the earliest cortical accumulator,
a study of cognitively normal adults found that the right
medial frontal cortex accumulates amyloid the earliest,
closely followed by bilateral precuneus
based on cross-sectional voxelwise analyses~\citep{Villeneuve2015}.
We presented regional trajectories averaged bilaterally in this work;
however, hemisphere-specific trajectories were consistent with our 
bilaterally-averaged trajectories and we did not observe that medial frontal 
cortex precedes precuneus.
Further studies with larger samples will be instrumental in elucidating the 
regional progression of amyloid accumulation.

One of the strengths of our model for the progression scores $s$ is that there are no 
\emph{a priori}
assumptions on the global form of $s(t)$, which can be thought of as an underlying function that
describes the evolution of progression score over time.
Instead, the model makes linear local approximations to $s(t)$ per individual.
The linear form we use makes the EM approach analytically tractable and 
enables the discovery of the global form of $s(t)$.
On the other hand, the linear relationship we 
assume between $s$ and $t$ is also a limitation since it may
be an oversimplification over long follow-up durations.
In order to capture dynamics over longer periods more accurately,
it is necessary to investigate the relationship between the progression
scores $s$ and time $t$, and select an appropriate function
to link these variables.

Another limitation of our model is the assumption of linear biomarker trajectories. 
This assumption is not reflective of the fact that voxelwise DVR has a theoretical minimum of 1 
(while in practice, there are DVR values lower than 1 due to noise). 
Furthermore, several studies of longitudinal amyloid deposition have found evidence for a ceiling 
effect in amyloid deposition in late AD, suggesting a sigmoid trajectory for amyloid 
levels~\citep{Villemagne2013,Villain2012}.
Our use of a linear rather than sigmoid trajectory inevitably results in inaccuracies in PS 
calculation for individuals whose voxelwise data lie along the plateaus of the sigmoid in reality. 
The linear trajectory assumption may also prevent us from characterizing subtler differences in 
the temporal ordering of the onset of amyloid accumulation across different regions. 
However, linear trajectories, unlike sigmoid trajectories, yield closed-form update equations for 
trajectory parameters, greatly facilitating the model fitting procedure.

The spatial covariance functions we used in our analyses are covariance functions of strictly 
stationary isotropic processes. 
The underlying spatial noise process is affected by the PET scanner, 
image reconstruction algorithm, radiotracer delivery and binding in different brain regions, 
kinetic parameter estimation algorithm, and registration methods used to bring all scans in 
alignment. 
To accurately model these influences on noise, complex noise spatial covariance models are needed. 
In this work, we made simplifying assumptions on the noise properties to facilitate the study of 
longitudinal trajectories of amyloid images. 
As a result of these simplifications, our model does 
not accurately capture noise properties; however, by eliminating the inaccurate assumption of 
independence across voxels, our approach improves the accuracy of the estimated trajectories and 
progression scores. 
Refined spatial noise models incorporating non-stationarity and anisotropy may further improve the 
estimation accuracy.

Our model can be applied to studying higher resolution images. 
To reduce computational memory burden, it may be necessary to impose sparsity on the spatial 
correlation matrix, which can be done by imposing a correlation value of 0 at large distances. 
The sparsity property of the correlation matrix can be taken advantage of to yield Cholesky 
decompositions using less memory and time.

In conclusion, the progression score model allows for extracting summary 
scores from longitudinal data for each scan.
In this work, we have extended the progression score model proposed 
by~\citet{Jedynak2012} to voxelwise imaging data
by accounting for spatial correlations and enabling efficient handling of the large
number of voxels through the EM framework.
The incorporation of a prior on subject-specific variables allowed for the 
inclusion of individuals with a single visit in the model.
Our method can be extended to the analysis of other types of imaging data, and 
in cases where a summary score such as mean cortical DVR is not available, the 
progression score estimates can be highly informative for progression staging.

\section{Acknowledgment}
We thank Kalyani Kansal for her thorough review of the mathematical derivations and
assistance with code testing.
This research was supported in part by the Intramural Research Program of
the National Institutes of Health and by the Michael J. Fox Foundation for Parkinson's
Research, MJFF Research Grant ID: 9310.

\appendix

\setcounter{figure}{0}
\setcounter{table}{0}

\setcounter{proposition}{0}
\renewcommand{\theproposition}{\Alph{section}.\arabic{proposition}}

\section{}
\label{app:EM}

\begin{proposition}
  \begin{eqnarray}
  f(\tilde{\mathbf{u}}_i \mid \mathbf{y}_i ; \boldsymbol \theta')
  & \propto &
  \Phi(\tilde{\mathbf{u}}_i; \hat{\mathbf{u}}'_i, \Sigma'_i) ,
\end{eqnarray}
where
$\Phi(\, \cdot \,; \boldsymbol \mu, \Sigma)$ denotes the probability density function of a 
multivariate normal with mean $\boldsymbol \mu$ and covariance matrix $\Sigma$,
\begin{equation}
\hat{\mathbf{u}}'_i = 
\left( \sum_j  Z'^T_{ij} R'^{-1} Z'_{ij} + V'^{-1} \right)^{-1}
\left( \sum_{j} Z'^T_{ij} R'^{-1}(\mathbf{y}_{ij} - \mathbf{b}') + V'^{-1} \mathbf{m}' 
\right),
\end{equation}
and
$\Sigma'_i = \left( \sum_j  Z'^T_{ij} R'^{-1} Z'_{ij} + V'^{-1} \right)^{-1}$ is a covariance matrix.
\label{lemma:uhat2}
\end{proposition}

\begin{proof}
Independence across visits allows us to write
\begin{eqnarray}
  f(\mathbf{y}_i \mid \tilde{\mathbf{u}}_i; \boldsymbol\theta')
  & = &
  \prod_j f(\mathbf{y}_{ij} \mid \tilde{\mathbf{u}}_i; \boldsymbol\theta') \nonumber \\
  &=&
  \prod_j \Phi(\mathbf{y}_{ij}; Z'_{ij} \tilde{\mathbf{u}}_i + \mathbf{b}', R')  
  \nonumber
  \\
  & \propto &
  \exp \left\{ -\frac{1}{2} 
  \left[ \tilde{\mathbf{u}}^T_i \left( \sum_j  Z'^T_{ij} R'^{-1} Z'_{ij} \right) 
  \tilde{\mathbf{u}}_i
  \right.
  \right.
  \nonumber
  \\
  &\hphantom{\propto}&
  \left.
  \left.
  - 2 \left( \sum_j (\mathbf{y}_{ij} - \mathbf{b}')^T R'^{-1} Z'_{ij} \right) 
  \tilde{\mathbf{u}}_i
  \right]
  \right\},
\end{eqnarray}
where we have ignored the terms that do not depend on $\tilde{\mathbf{u}}_i$.

By Bayes' rule,
\begin{eqnarray}
  f(\tilde{\mathbf{u}}_i \mid \mathbf{y}_i; \boldsymbol\theta')
  & \propto &
  f(\mathbf{y}_i \mid \tilde{\mathbf{u}}_i; \boldsymbol\theta')
  f(\tilde{\mathbf{u}}_i; \boldsymbol\theta')
  \\
  & \propto &
  \exp \left\{
  -\frac{1}{2} \left[ 
  \tilde{\mathbf{u}}^T_i \left( \sum_j  Z'^T_{ij} R'^{-1} Z'_{ij} + V'^{-1} \right) \tilde{\mathbf{u}}_i
  \right.
  \right.
  \nonumber
  \\
  &\hphantom{\propto}&
  \left.
  \left.
  -2 \left( \sum_j (\mathbf{y}_{ij} - \mathbf{b}')^T R'^{-1} Z'_{ij} + \mathbf{m}'^T V'^{-1} \right)
  \tilde{\mathbf{u}}_i
  \right]
  \right\}
  \\
  & \propto &
  \Phi(\tilde{\mathbf{u}}_i; \hat{\mathbf{u}}_i, \Sigma'_i)
\end{eqnarray}

\end{proof}

Below we derive the EM algorithm update equations given in 
Section~\ref{sec:EM-update-eq}:

\subsection*{Solving for the intercept parameter $\mathbf{b}$}
The value of $\mathbf{b}$ that solves
$\frac{\partial Q}{\partial \mathbf{b}}=R^{-1} \sum_{i,j} (\mathbf{y}_{ij}-Z_{ij} \hat{\mathbf{u}}'_i-\mathbf{b})=0$ is given by
\begin{eqnarray}
  \mathbf{b} & = & \frac{1}{\sum_i v_i}
  \sum_{i,j} 
  \left(
  \mathbf{y}_{ij} - Z_{ij} \hat{\mathbf{u}}'_i
  \right) \nonumber
  \\
  & = &
  \frac{1}{\sum_i v_i}
  \sum_{i,j} 
  \left(
  \mathbf{y}_{ij} - \mathbf{a} s'_{ij}
  \right).
  \label{eq:b-prelim}
\end{eqnarray}

Plugging in the expression for $\mathbf{a}$ from Equation~\ref{eq:a-prelim} yields
\begin{equation}
  \mathbf{b}
  =
  \frac{
  \left(
  \sum_{i,j}\mathbf{y}_{ij}
  \right)
  \left(
  \sum_{i,j} \mathbf{q}^T_{ij} \Sigma'_i \mathbf{q}_{ij} +
  s'^2_{ij}
  \right)
  -
  \left(
  \sum_{i,j} \mathbf{y}_{ij} s'_{ij}
  \right)
  \left(
  \sum_{i,j} s'_{ij}
  \right)
  }
  {
  \left(
  \sum_i v_i
  \right)
  \left(
  \sum_{i,j} \mathbf{q}^T_{ij} \Sigma'_i \mathbf{q}_{ij} +
  s'^2_{ij}
  \right)
  -
  \left(
  \sum_{i,j} s'_{ij}
  \right)^2
  }.
  \label{eq:b}
\end{equation}

\subsection*{Solving for the slope parameter $\mathbf{a}$}
The value of $\mathbf{a}$ that solves 
  ${\frac{\partial Q}{\partial \mathbf{a}} = 
  R^{-1} 
  \left[
  \sum_{i,j} (\mathbf{y}_{ij}-\mathbf{b}) s'_{ij}
  - \mathbf{a}
  \left(
  \sum_{i,j} \mathbf{q}^T_{ij} \Sigma'_i \mathbf{q}_{ij} +
  s'^2_{ij}
  \right) 
  \right] = 0}$ is given 
  by
\begin{equation}
  \mathbf{a} = 
  \frac{1}
  {\sum_{i,j} \mathbf{q}^T_{ij} \Sigma'_i \mathbf{q}_{ij} +
  s'^2_{ij}}
  \sum_{i,j} (\mathbf{y}_{ij}-\mathbf{b}) s'_{ij}.
  \label{eq:a-prelim}
\end{equation}

Plugging in the expression for $\mathbf{b}$ from Equation~\ref{eq:b-prelim} yields
\begin{eqnarray}
  \mathbf{a} &=& 
  \frac{ \left(\sum_i v_i \right) \left( \sum_{i,j} \mathbf{y}_{ij} s'_{ij} \right) -
  \left( \sum_{i,j} \mathbf{y}_{ij} \right) \left( \sum_{i,j} s'_{ij} \right) }
  { \left(\sum_i v_i \right)
  \left( \sum_{i,j} \mathbf{q}^T_{ij} \Sigma'_i \mathbf{q}_{ij} +
  s'^2_{ij} \right) - \left( \sum_{i,j} s'_{ij} \right)^2
  }.
  \label{eq:a}
\end{eqnarray}

\subsection*{Solving for the subject-specific variable mean parameter $\mathbf{m}$}

The value of $\mathbf{m}$ that solves 
$\frac{\partial Q}{\partial \mathbf{m}} =-\sum_{i} V^{-1} \mathbf{m} + \sum_{i} V^{-1} \hat{\mathbf{u}}'_i=0$
is given by
\begin{eqnarray}
  \mathbf{m} = \frac{1}{n} \sum_i \hat{\mathbf{u}}'_i.
\end{eqnarray}

\subsection*{Solving for the subject-specific variable variance parameter $\boldsymbol \nu$}
We use numerical optimization to estimate $\boldsymbol \nu$:
\begin{eqnarray}
  \boldsymbol \nu &=& \argmax_{\boldsymbol \nu} Q(\boldsymbol \theta, \boldsymbol \theta')
  \\
  &=&
  \argmin_{\boldsymbol \nu}
  \sum_{i} \left( \log |V|
  +(\hat{\mathbf{u}}'_i - \mathbf{m})^T V^{-1} (\hat{\mathbf{u}}'_i - \mathbf{m})
  +\text{Tr} \left( V^{-1} \Sigma'_i \right) \right).
\end{eqnarray}
We can reconstruct $V$ by undoing the log-Cholesky parametrization steps using the
estimated parameter vector $\boldsymbol \nu = [\nu_1, \nu_2, \nu_3]^T$ as
\begin{eqnarray}
  V &=&
  \begin{bmatrix}
    e^{\nu_1} & 0 \\
    \nu_2 & e^{\nu_3}
  \end{bmatrix}
    \begin{bmatrix}
    e^{\nu_1} & \nu_2 \\
    0 & e^{\nu_3}
  \end{bmatrix}.
\end{eqnarray}

\subsection*{Solving for the noise covariance parameters $\boldsymbol\lambda$ and $\rho$}
If $C = I_{K \times K}$ (i.e. $\rho = 0$ and fixed), then 
the diagonal elements $\lambda_k$ of $\Lambda$ can be estimated as:
\begin{eqnarray}
  \lambda_k &=& \sqrt{\frac{1}{\sum_i v_i} 
  \sum_{i,j} \left[ (y_{ijk} - a_k s'_{ij} - b_k)^2
  + a^2_k \mathbf{q}^T_{ij} \Sigma'_i \mathbf{q}_{ij}
  \right]}.
  \label{eq:multiLambda}
\end{eqnarray}

If $C$ is not fixed to be the identity matrix, then in general it is not possible to obtain 
closed-form solutions for $\boldsymbol\lambda$ and $\rho$
and we must use numerical optimization over $Q$.
When $\boldsymbol\lambda$ is a high-dimensional vector, this is not feasible.
Therefore,
we simplify the parametrization of the noise covariance matrix $R$ as
$R(\lambda,\rho)=\lambda^2 \hat{\Lambda} C(\rho) \hat{\Lambda}$, where
we now consider $\hat{\Lambda}$
as a fixed diagonal matrix with positive diagonal entries.
We fix $\hat{\Lambda}$ at the estimate of $\Lambda$ 
from the model with $C=I_{K \times K}$.
$\rho$ is the correlation matrix parameter as defined previously, 
and $\lambda > 0$ is a scaling parameter.
Now we need to perform numerical optimization over $Q$ to estimate only two 
parameters:
\begin{eqnarray}
  \lambda, \rho = \argmax_{\lambda,\rho} Q(\boldsymbol \theta, \boldsymbol 
  \theta').
\end{eqnarray}
Note that since the EM algorithm does not require the maximization of
$Q(\boldsymbol \theta,\boldsymbol \theta')$ but simply requires an increase in this 
function with each iteration, performing numerical optimization until convergence is 
not necessary.

\begin{proposition}
  Consider
the model given in Section~\ref{sec:overall-model} with
$\boldsymbol\theta = \{\mathbf{m}, V, \mathbf{a}, \mathbf{b}, R \}$.
The reparametrization
$\boldsymbol\theta^* = \left\{w\mathbf{m}+
\begin{bmatrix}
  0 \\ z
\end{bmatrix}, w^2 V, \frac{1}{w} \mathbf{a}, \mathbf{b} - \frac{z}{w} \mathbf{a}, R 
\right\}$, where $w,z \in \mathbb{R}$, $w \neq 0$ yields an equivalent model.
\label{lemma:param-standardization}
\end{proposition}

\begin{proof}
  The incomplete log-likelihood is
  \begin{eqnarray}
    \log f(\mathbf{y}; \boldsymbol\theta) &=&
    \frac{1}{2} \sum_i \left( \log | 2 \pi \Sigma_i |
    - \log | 2 \pi V | - v_i \log | 2 \pi R | \right) \nonumber \\
    &\hphantom{=}&
    -\frac{1}{2} \sum_{i,j} (\mathbf{y}_{ij} -\mathbf{b})^T R^{-1} (\mathbf{y}_{ij} -\mathbf{b}) 
    \nonumber \\
    &\hphantom{=}&
    -\frac{1}{2} \sum_i \mathbf{m}^T V^{-1} \mathbf{m}
    +\frac{1}{2} \sum_i \hat{\mathbf{u}}_i^T \Sigma^{-1}_i \hat{\mathbf{u}}_i.
  \end{eqnarray}
  Here, we consider the case where $z=0$ for algebraic simplicity.
  For the reparameterized model, we obtain
    $\hat{\mathbf{u}}_i^* = w \hat{\mathbf{u}}_i$ and
    $\Sigma^*_i = w^2 \Sigma_i$.
  Therefore, each of the terms in $\log f(\mathbf{y}; \boldsymbol\theta^*)$ remain 
  the same as those in $\log f(\mathbf{y}; \boldsymbol\theta)$, yielding
  $\log f(\mathbf{y}; \boldsymbol\theta^*) = \log f(\mathbf{y}; 
  \boldsymbol\theta)$.
  \end{proof}

\clearpage

\beginsupplement

\begin{figure}
  \centering
  \includegraphics[width=.7\textwidth]{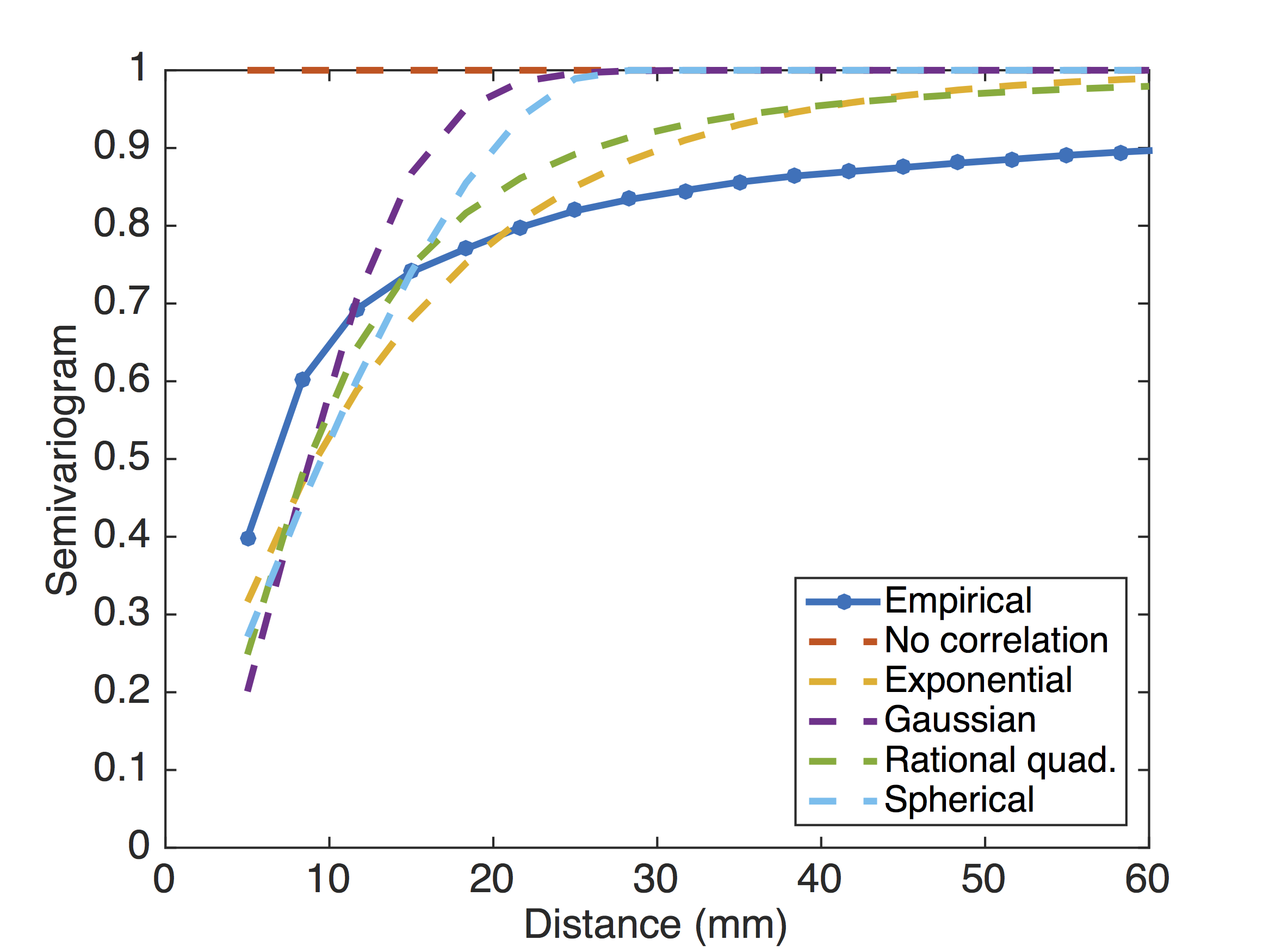}
  \caption{Preliminary analysis of the noise spatial correlation structure using 
  the semivariogram.
  The empirical semivariogram computed using the residuals from the model where
  $C=I_{K \times K}$ is shown in blue.
  We fitted the semivariograms corresponding to the exponential, Gaussian, 
  rational quadratic, and spherical correlation structures listed in 
  Table~\ref{tab:corr-fun} to the empirical semivariogram.
  The rational quadratic function had the best fit to the empirical curve
  based on the sum of squared error of the fitted semivariogram over a 100~mm distance
  calculated using 30~equidistant points.}
  \label{fig:semivariogram}
\end{figure}

\begin{figure}
  \centering
  \includegraphics[width=1\textwidth]{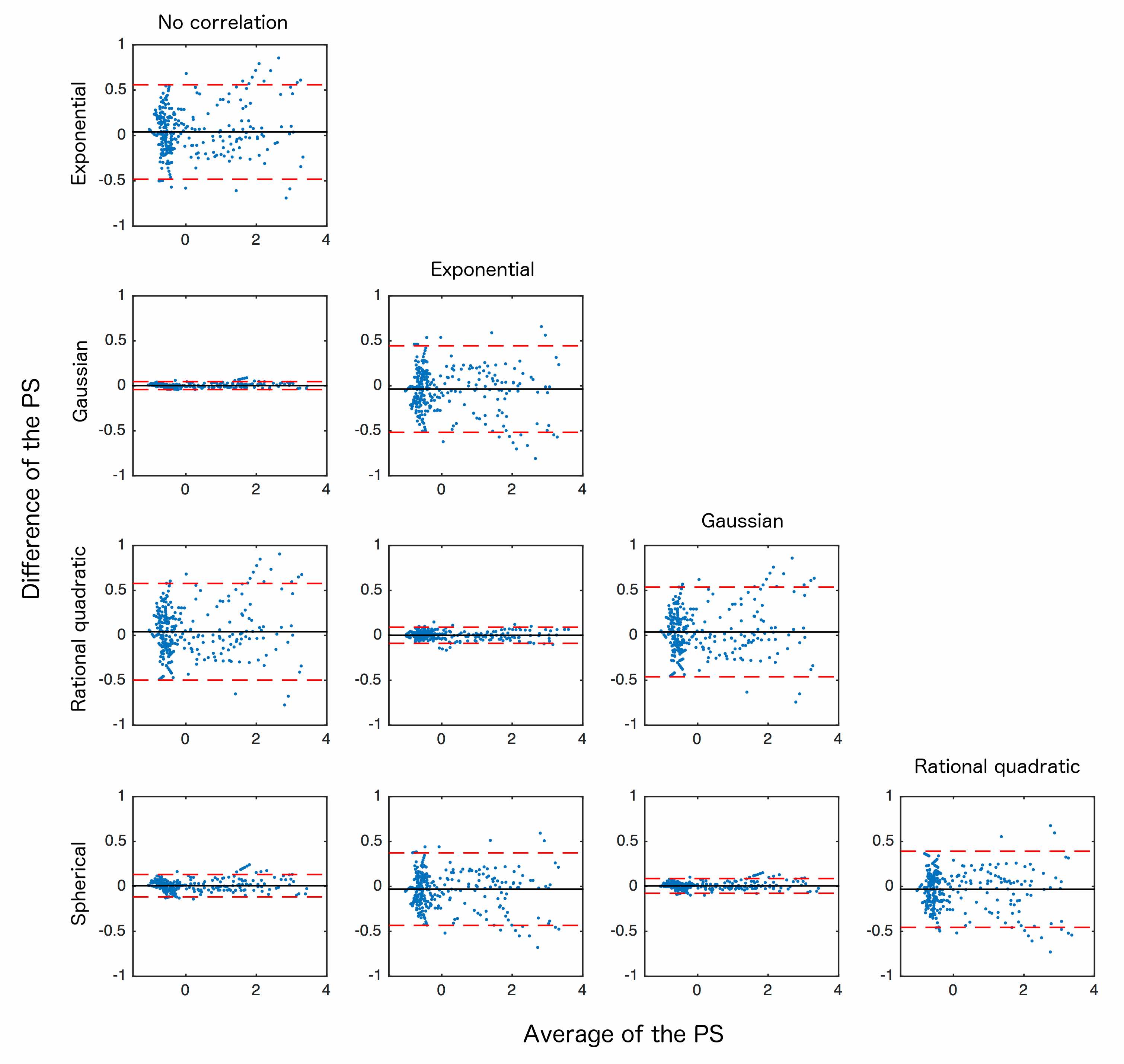}
  \caption{Bland-Altman plots comparing A$\beta$-PS values obtained using different 
  spatial correlation models.
  We fitted the model using the ``no correlation'' structure ($C=I_{K \times K}$) 
  and four spatial correlation structures (exponential, Gaussian, rational quadratic,
  and spherical).
  We then used Bland-Altman plots to assess the differences in PS computed using 
  these correlation structures.
  In the Bland-Altman plots, the $x$-axis is the average of the PS values computed 
  using the two methods being compared, and the $y$-axis is the difference.
  The labels above and to the left of the figures describe the spatial 
  correlation structures being compared.
  Each blue dot corresponds to a visit.
  The solid black line indicates the mean difference between PS values computed 
  using the methods, and the dashed red lines indicate its 95\% confidence band.
  The Bland-Altman plots indicate that the ``no correlation'', Gaussian, and 
  spherical structures yield similar PS values.
  Furthermore, the exponential and rational quadratic structures also yield 
  similar PS values, but these differ from those obtained using the other three 
  structures.}
  \label{fig:bland-altman-ps}
\end{figure}

\begin{figure}
  \centering
  \includegraphics[width=1\textwidth]{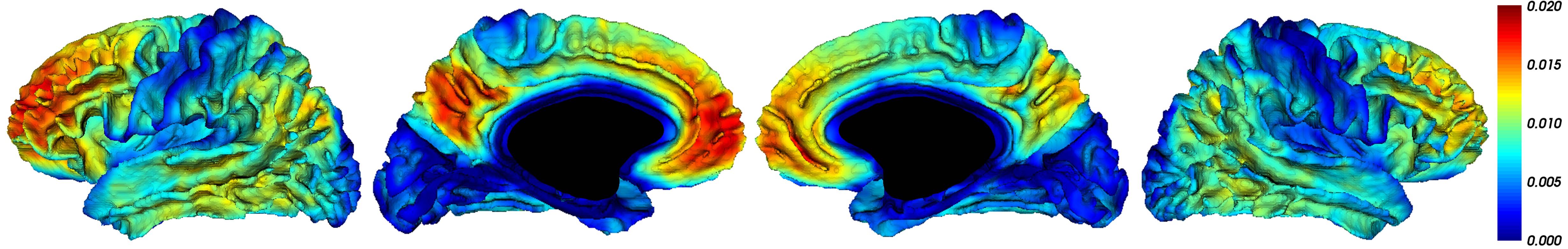}
  \caption{Slope parameters $\eta_k$ (fixed effects) obtained from voxelwise LME model projected
  onto the cortical surface. DVR value at voxel $k$ increases by $\eta_k$ per year on average.}
  \label{fig:lme-A}
\end{figure}

\begin{figure}
  \centering
  \includegraphics[width=.7\textwidth]{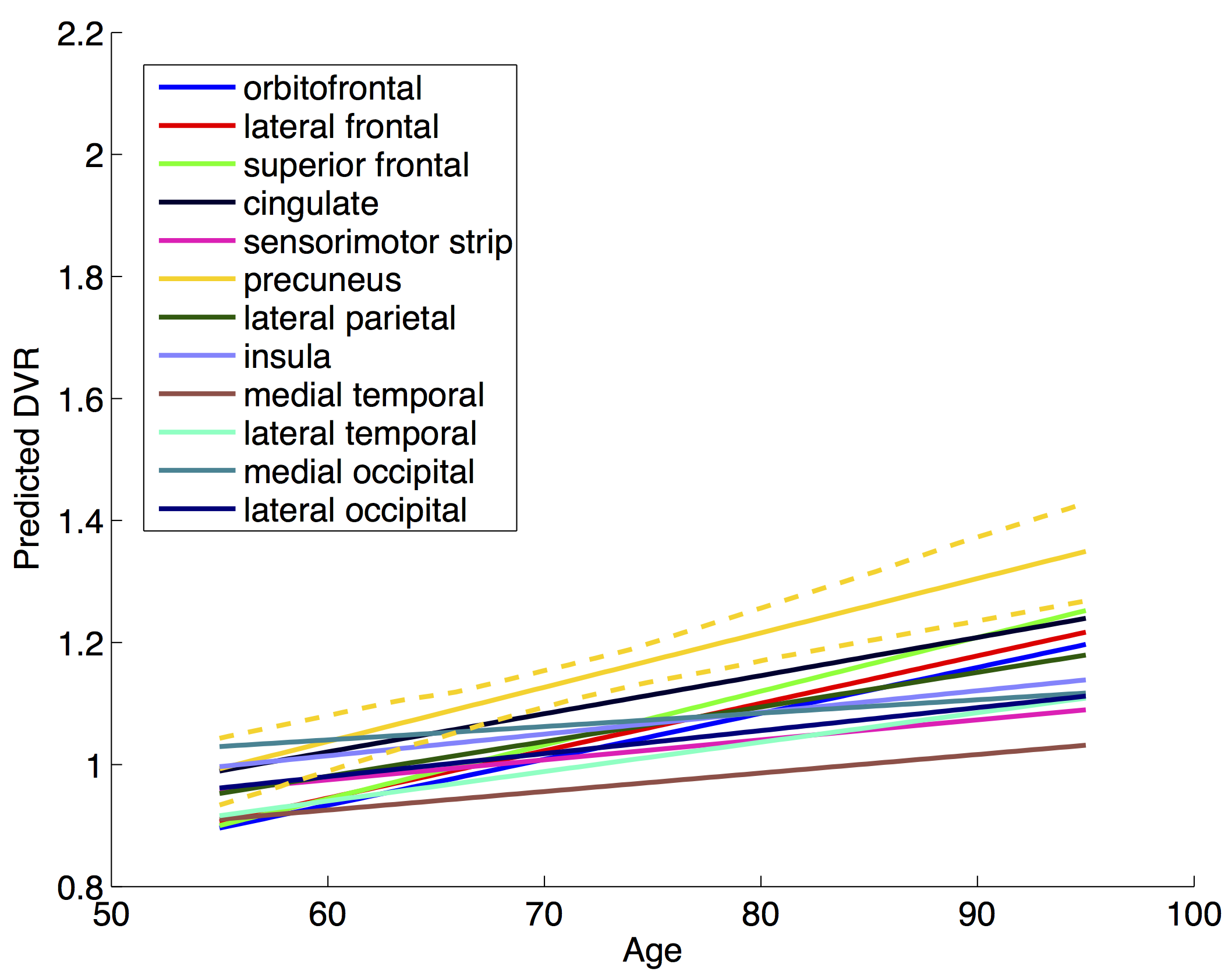}
  \caption{Regional trajectories obtained from the LME model as function of age.
  The LME model was used to make voxelwise predictions at a range of
  age values based on the estimated fixed effects,
  and these predictions were averaged within each ROI to obtain regional
  trajectories.
  The dashed lines indicate the 95\% confidence bands for the cortical regions.}
  \label{fig:precuneus-trajectory-CI-lme}
\end{figure}

\begin{figure}
  \centering
  \includegraphics[width=1\textwidth]{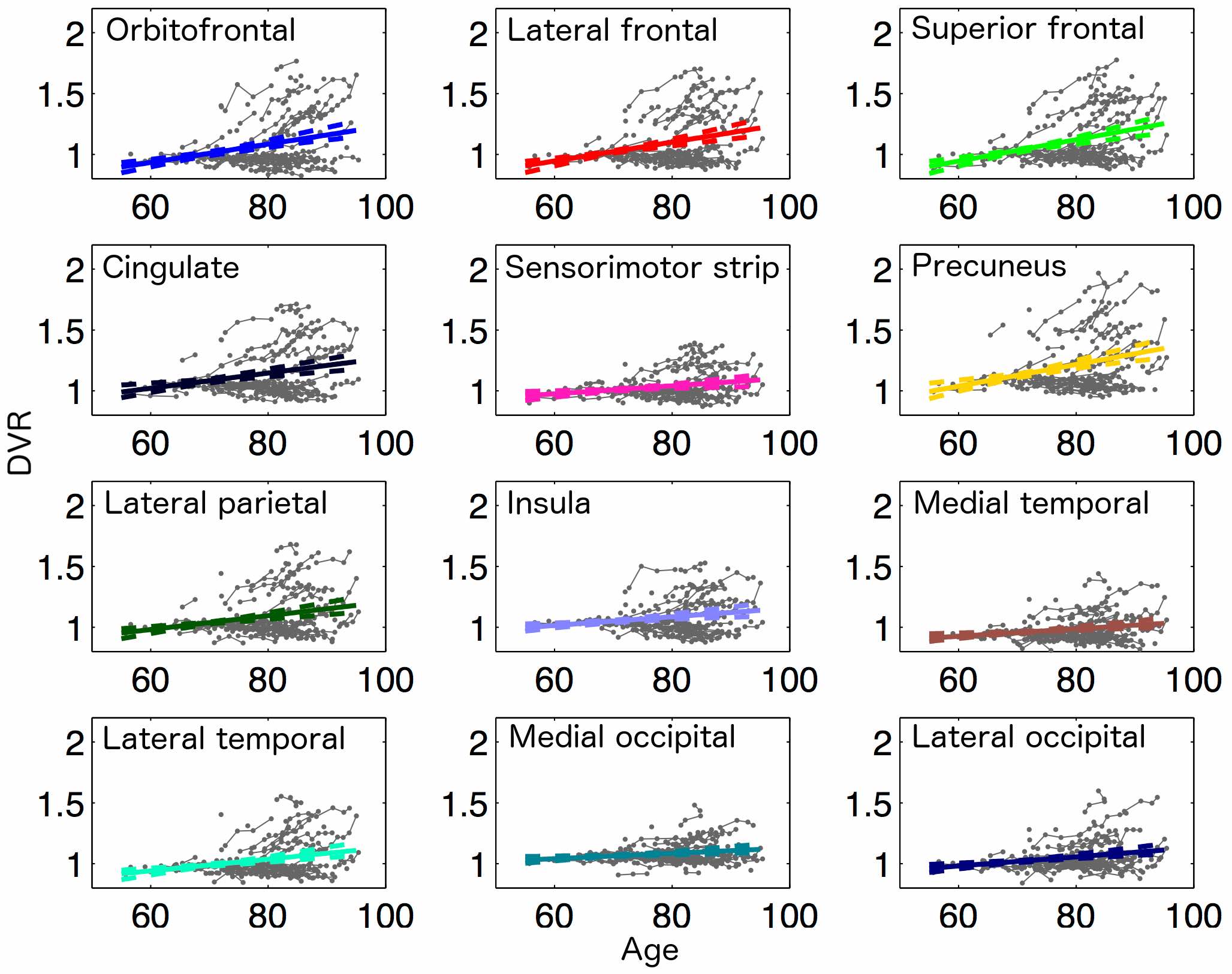}
  \caption{Regional trajectories obtained from the LME model as function of age.
  The LME model was used to make voxelwise predictions at a range of
  age values based on the estimated fixed effects, 
  and these predictions were averaged within each ROI to obtain regional
  trajectories.
  The dashed lines indicate the 95\% confidence bands for the cortical regions.
  Estimated trajectories with their 95\% confidence bands are superimposed on 
  observed longitudinal data (in gray).}
  \label{fig:ROI-trajectory-CI-lme}
\end{figure}

\begin{figure}
  \centering
  \includegraphics[width=1\textwidth]{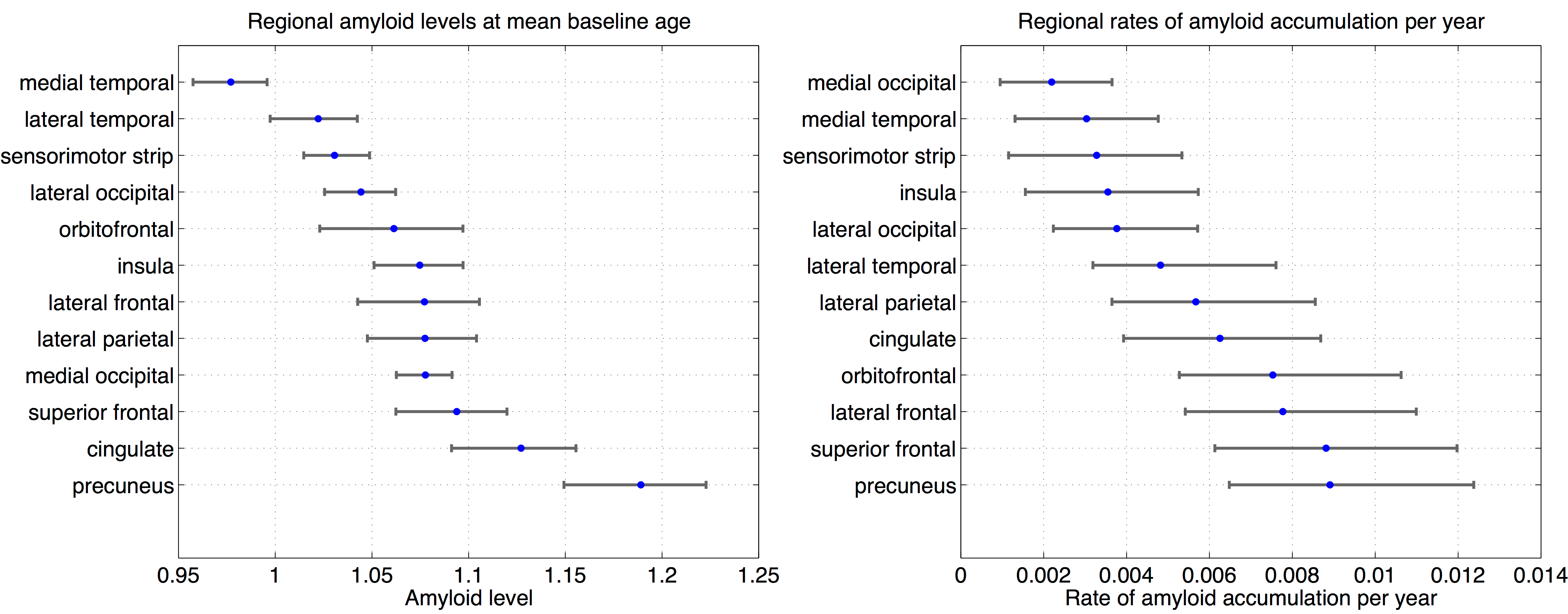}
  \caption{Comparison of levels of amyloid at age 77 (the mean baseline age of the 
  sample) and rates of amyloid accumulation across cortical regions
  using results of the LME model.
  We used the fixed effect estimates to obtain voxelwise amyloid levels at age 77
  as $\hat{y}_k = \eta_k \times 77 + \gamma_k$, and averaged $\hat{y}_k$ within 
  each ROI to obtain regional amyloid levels at age 77.
  The fixed effects $\eta_k$ were averaged within each ROI to obtain regional
  rates.}
  \label{fig:comparison-ROI-lme}
\end{figure}

\clearpage

\section*{References}

\bibliography{Bilgel-Amyloid_PS-arxiv}

\begin{thebibliography}{46}
\expandafter\ifx\csname natexlab\endcsname\relax\def\natexlab#1{#1}\fi
\providecommand{\url}[1]{\texttt{#1}}
\providecommand{\href}[2]{#2}
\providecommand{\path}[1]{#1}
\providecommand{\DOIprefix}{doi:}
\providecommand{\ArXivprefix}{arXiv:}
\providecommand{\URLprefix}{URL: }
\providecommand{\Pubmedprefix}{pmid:}
\providecommand{\doi}[1]{\href{http://dx.doi.org/#1}{\path{#1}}}
\providecommand{\Pubmed}[1]{\href{pmid:#1}{\path{#1}}}
\providecommand{\bibinfo}[2]{#2}
\ifx\xfnm\relax \def\xfnm[#1]{\unskip,\space#1}\fi
\bibitem[{Avants et~al.(2008)Avants, Epstein, Grossman and Gee}]{Avants2008}
\bibinfo{author}{Avants, B.B.}, \bibinfo{author}{Epstein, C.L.},
  \bibinfo{author}{Grossman, M.}, \bibinfo{author}{Gee, J.C.},
  \bibinfo{year}{2008}.
\newblock \bibinfo{title}{{Symmetric diffeomorphic image registration with
  cross-correlation: evaluating automated labeling of elderly and
  neurodegenerative brain}}.
\newblock \bibinfo{journal}{Medical Image Analysis} \bibinfo{volume}{12},
  \bibinfo{pages}{26--41}.
\bibitem[{Avants et~al.(2010)Avants, Yushkevich, Pluta, Minkoff, Korczykowski,
  Detre and Gee}]{Avants2010}
\bibinfo{author}{Avants, B.B.}, \bibinfo{author}{Yushkevich, P.},
  \bibinfo{author}{Pluta, J.}, \bibinfo{author}{Minkoff, D.},
  \bibinfo{author}{Korczykowski, M.}, \bibinfo{author}{Detre, J.},
  \bibinfo{author}{Gee, J.C.}, \bibinfo{year}{2010}.
\newblock \bibinfo{title}{{The optimal template effect in hippocampus studies
  of diseased populations}}.
\newblock \bibinfo{journal}{NeuroImage} \bibinfo{volume}{49},
  \bibinfo{pages}{2457--2466}.
\bibitem[{Bateman et~al.(2012)Bateman, Xiong, Benzinger, Fagan, Goate, Fox,
  Marcus, Cairns, Xie, Blazey, Holtzman, Santacruz, Buckles, Oliver, Moulder,
  Aisen, Ghetti, Klunk, McDade, Martins, Masters, Mayeux, Ringman, Rossor,
  Schofield, Sperling, Salloway and Morris}]{Bateman2012}
\bibinfo{author}{Bateman, R.J.}, \bibinfo{author}{Xiong, C.},
  \bibinfo{author}{Benzinger, T.L.S.}, \bibinfo{author}{Fagan, A.M.},
  \bibinfo{author}{Goate, A.}, \bibinfo{author}{Fox, N.C.},
  \bibinfo{author}{Marcus, D.S.}, \bibinfo{author}{Cairns, N.J.},
  \bibinfo{author}{Xie, X.}, \bibinfo{author}{Blazey, T.M.},
  \bibinfo{author}{Holtzman, D.M.}, \bibinfo{author}{Santacruz, A.},
  \bibinfo{author}{Buckles, V.}, \bibinfo{author}{Oliver, A.},
  \bibinfo{author}{Moulder, K.}, \bibinfo{author}{Aisen, P.S.},
  \bibinfo{author}{Ghetti, B.}, \bibinfo{author}{Klunk, W.E.},
  \bibinfo{author}{McDade, E.}, \bibinfo{author}{Martins, R.N.},
  \bibinfo{author}{Masters, C.L.}, \bibinfo{author}{Mayeux, R.},
  \bibinfo{author}{Ringman, J.M.}, \bibinfo{author}{Rossor, M.N.},
  \bibinfo{author}{Schofield, P.R.}, \bibinfo{author}{Sperling, R.A.},
  \bibinfo{author}{Salloway, S.}, \bibinfo{author}{Morris, J.C.},
  \bibinfo{year}{2012}.
\newblock \bibinfo{title}{{Clinical and biomarker changes in dominantly
  inherited Alzheimer's disease}}.
\newblock \bibinfo{journal}{New England Journal of Medicine}
  \bibinfo{volume}{367}, \bibinfo{pages}{795--804}.
\bibitem[{Bernal-Rusiel et~al.(2012)Bernal-Rusiel, Greve, Reuter, Fischl and
  Sabuncu}]{Bernal-Rusiel2012}
\bibinfo{author}{Bernal-Rusiel, J.L.}, \bibinfo{author}{Greve, D.N.},
  \bibinfo{author}{Reuter, M.}, \bibinfo{author}{Fischl, B.},
  \bibinfo{author}{Sabuncu, M.R.}, \bibinfo{year}{2012}.
\newblock \bibinfo{title}{{Statistical analysis of longitudinal neuroimage data
  with linear mixed effects models}}.
\newblock \bibinfo{journal}{NeuroImage} \bibinfo{volume}{66},
  \bibinfo{pages}{249--260}.
\bibitem[{Bernal-Rusiel et~al.(2013)Bernal-Rusiel, Reuter, Greve, Fischl and
  Sabuncu}]{Bernal-Rusiel2013}
\bibinfo{author}{Bernal-Rusiel, J.L.}, \bibinfo{author}{Reuter, M.},
  \bibinfo{author}{Greve, D.N.}, \bibinfo{author}{Fischl, B.},
  \bibinfo{author}{Sabuncu, M.R.}, \bibinfo{year}{2013}.
\newblock \bibinfo{title}{{Spatiotemporal linear mixed effects modeling for the
  mass-univariate analysis of longitudinal neuroimage data}}.
\newblock \bibinfo{journal}{NeuroImage} \bibinfo{volume}{81},
  \bibinfo{pages}{358--370}.
\bibitem[{Bilgel et~al.(2014)Bilgel, An, Lang, Prince, Ferrucci, Jedynak and
  Resnick}]{Bilgel2014b}
\bibinfo{author}{Bilgel, M.}, \bibinfo{author}{An, Y.}, \bibinfo{author}{Lang,
  A.}, \bibinfo{author}{Prince, J.}, \bibinfo{author}{Ferrucci, L.},
  \bibinfo{author}{Jedynak, B.}, \bibinfo{author}{Resnick, S.M.},
  \bibinfo{year}{2014}.
\newblock \bibinfo{title}{{Trajectories of Alzheimer disease-related cognitive
  measures in a longitudinal sample.}}
\newblock \bibinfo{journal}{Alzheimer's \& Dementia} \bibinfo{volume}{10},
  \bibinfo{pages}{735--742}.
\bibitem[{Bilgel et~al.(2015a)Bilgel, An, Zhou, Wong, Prince, Ferrucci and
  Resnick}]{Bilgel2015b}
\bibinfo{author}{Bilgel, M.}, \bibinfo{author}{An, Y.}, \bibinfo{author}{Zhou,
  Y.}, \bibinfo{author}{Wong, D.F.}, \bibinfo{author}{Prince, J.L.},
  \bibinfo{author}{Ferrucci, L.}, \bibinfo{author}{Resnick, S.M.},
  \bibinfo{year}{2015}a.
\newblock \bibinfo{title}{{Individual estimates of age at detectable amyloid
  onset for risk factor assessment}}.
\newblock \bibinfo{journal}{Alzheimer's {\&} Dementia} \bibinfo{volume}{[In
  press]}.
\bibitem[{Bilgel et~al.(2015b)Bilgel, Carass, Resnick, Wong and
  Prince}]{Bilgel2015}
\bibinfo{author}{Bilgel, M.}, \bibinfo{author}{Carass, A.},
  \bibinfo{author}{Resnick, S.M.}, \bibinfo{author}{Wong, D.F.},
  \bibinfo{author}{Prince, J.L.}, \bibinfo{year}{2015}b.
\newblock \bibinfo{title}{{Deformation field correction for spatial
  normalization of PET images}}.
\newblock \bibinfo{journal}{NeuroImage} \bibinfo{volume}{119},
  \bibinfo{pages}{152--163}.
\bibitem[{Bilgel et~al.(2015c)Bilgel, Jedynak, Wong, Resnick and
  Prince}]{Bilgel2015a}
\bibinfo{author}{Bilgel, M.}, \bibinfo{author}{Jedynak, B.},
  \bibinfo{author}{Wong, D.F.}, \bibinfo{author}{Resnick, S.M.},
  \bibinfo{author}{Prince, J.L.}, \bibinfo{year}{2015}c.
\newblock \bibinfo{title}{{Temporal trajectory and progression score estimation
  from voxelwise longitudinal imaging measures: Application to amyloid
  imaging}}, in: \bibinfo{editor}{Ourselin, S.}, \bibinfo{editor}{Alexander,
  D.C.}, \bibinfo{editor}{Westin, C.F.}, \bibinfo{editor}{Cardoso, M.J.}
  (Eds.), \bibinfo{booktitle}{Lecture Notes in Computer Science (9123),
  Information Processing in Medical Imaging}, pp. \bibinfo{pages}{424--436}.
\bibitem[{{Biomarkers Definitions Working Group}(2001)}]{Definitions2001}
\bibinfo{author}{{Biomarkers Definitions Working Group}}, \bibinfo{year}{2001}.
\newblock \bibinfo{title}{{Biomarkers and surrogate endpoints: Preferred
  definitions and conceptual framework}}.
\newblock \bibinfo{journal}{Clinical Pharmacology {\&} Therapeutics}
  \bibinfo{volume}{69}, \bibinfo{pages}{89--95}.
\bibitem[{Braak and Braak(1991)}]{Braak1991}
\bibinfo{author}{Braak, H.}, \bibinfo{author}{Braak, E.}, \bibinfo{year}{1991}.
\newblock \bibinfo{title}{{Neuropathological staging of Alzheimer-related
  changes}}.
\newblock \bibinfo{journal}{Acta Neuropathologica} \bibinfo{volume}{82},
  \bibinfo{pages}{239--259}.
\bibitem[{Caroli and Frisoni(2010)}]{Caroli2010}
\bibinfo{author}{Caroli, A.}, \bibinfo{author}{Frisoni, G.B.},
  \bibinfo{year}{2010}.
\newblock \bibinfo{title}{{The dynamics of Alzheimer's disease biomarkers in
  the Alzheimer's Disease Neuroimaging Initiative cohort}}.
\newblock \bibinfo{journal}{Neurobiology of Aging} \bibinfo{volume}{31},
  \bibinfo{pages}{1263--1274}.
\bibitem[{Cressie and Hawkins(1980)}]{Cressie1980}
\bibinfo{author}{Cressie, N.}, \bibinfo{author}{Hawkins, D.M.},
  \bibinfo{year}{1980}.
\newblock \bibinfo{title}{{Robust estimation of the variogram}}.
\newblock \bibinfo{journal}{Journal of the International Association of
  Mathematical Geology} \bibinfo{volume}{12}, \bibinfo{pages}{115--125}.
\bibitem[{Dale et~al.(1999)Dale, Fischl and Sereno}]{Dale1999}
\bibinfo{author}{Dale, A.}, \bibinfo{author}{Fischl, B.},
  \bibinfo{author}{Sereno, M.}, \bibinfo{year}{1999}.
\newblock \bibinfo{title}{{Cortical surface-based analysis: I. Segmentation and
  surface reconstruction}}.
\newblock \bibinfo{journal}{NeuroImage} \bibinfo{volume}{194},
  \bibinfo{pages}{179--194}.
\bibitem[{Delor et~al.(2013)Delor, Charoin, Gieschke, Retout and
  Jacqmin}]{Delor2013}
\bibinfo{author}{Delor, I.}, \bibinfo{author}{Charoin, J.E.},
  \bibinfo{author}{Gieschke, R.}, \bibinfo{author}{Retout, S.},
  \bibinfo{author}{Jacqmin, P.}, \bibinfo{year}{2013}.
\newblock \bibinfo{title}{{Modeling Alzheimer's disease progression using
  disease onset time and disease trajectory concepts applied to CDR-SOB scores
  from ADNI}}.
\newblock \bibinfo{journal}{CPT: Pharmacometrics {\&} Systems Pharmacology}
  \bibinfo{volume}{2}, \bibinfo{pages}{e78}.
\bibitem[{Desikan et~al.(2006)Desikan, S\'{e}gonne, Fischl, Quinn, Dickerson,
  Blacker, Buckner, Dale, Maguire, Hyman, Albert and Killiany}]{Desikan2006}
\bibinfo{author}{Desikan, R.S.}, \bibinfo{author}{S\'{e}gonne, F.},
  \bibinfo{author}{Fischl, B.}, \bibinfo{author}{Quinn, B.T.},
  \bibinfo{author}{Dickerson, B.C.}, \bibinfo{author}{Blacker, D.},
  \bibinfo{author}{Buckner, R.L.}, \bibinfo{author}{Dale, A.M.},
  \bibinfo{author}{Maguire, R.P.}, \bibinfo{author}{Hyman, B.T.},
  \bibinfo{author}{Albert, M.S.}, \bibinfo{author}{Killiany, R.J.},
  \bibinfo{year}{2006}.
\newblock \bibinfo{title}{{An automated labeling system for subdividing the
  human cerebral cortex on MRI scans into gyral based regions of interest}}.
\newblock \bibinfo{journal}{NeuroImage} \bibinfo{volume}{31},
  \bibinfo{pages}{968--980}.
\bibitem[{Donohue et~al.(2014)Donohue, Jacqmin-Gadda, {Le Goff}, Thomas, Raman,
  Gamst, Beckett, Jack, Weiner, Dartigues and Aisen}]{Donohue2014}
\bibinfo{author}{Donohue, M.C.}, \bibinfo{author}{Jacqmin-Gadda, H.},
  \bibinfo{author}{{Le Goff}, M.}, \bibinfo{author}{Thomas, R.G.},
  \bibinfo{author}{Raman, R.}, \bibinfo{author}{Gamst, A.C.},
  \bibinfo{author}{Beckett, L.A.}, \bibinfo{author}{Jack, C.R.},
  \bibinfo{author}{Weiner, M.W.}, \bibinfo{author}{Dartigues, J.F.},
  \bibinfo{author}{Aisen, P.S.}, \bibinfo{year}{2014}.
\newblock \bibinfo{title}{{Estimating long-term multivariate progression from
  short-term data}}.
\newblock \bibinfo{journal}{Alzheimer's and Dementia} \bibinfo{volume}{10},
  \bibinfo{pages}{S400--S410}.
\bibitem[{Doody et~al.(2010)Doody, Pavlik, Massman, Rountree, Darby and
  Chan}]{Doody2010}
\bibinfo{author}{Doody, R.S.}, \bibinfo{author}{Pavlik, V.},
  \bibinfo{author}{Massman, P.}, \bibinfo{author}{Rountree, S.},
  \bibinfo{author}{Darby, E.}, \bibinfo{author}{Chan, W.},
  \bibinfo{year}{2010}.
\newblock \bibinfo{title}{{Predicting progression of Alzheimer's disease}}.
\newblock \bibinfo{journal}{Alzheimer's Research {\&} Therapy}
  \bibinfo{volume}{2}.
\bibitem[{Fonteijn et~al.(2012)Fonteijn, Modat, Clarkson, Barnes, Lehmann,
  Hobbs, Scahill, Tabrizi, Ourselin, Fox and Alexander}]{Fonteijn2012}
\bibinfo{author}{Fonteijn, H.M.}, \bibinfo{author}{Modat, M.},
  \bibinfo{author}{Clarkson, M.J.}, \bibinfo{author}{Barnes, J.},
  \bibinfo{author}{Lehmann, M.}, \bibinfo{author}{Hobbs, N.Z.},
  \bibinfo{author}{Scahill, R.I.}, \bibinfo{author}{Tabrizi, S.J.},
  \bibinfo{author}{Ourselin, S.}, \bibinfo{author}{Fox, N.C.},
  \bibinfo{author}{Alexander, D.C.}, \bibinfo{year}{2012}.
\newblock \bibinfo{title}{{An event-based model for disease progression and its
  application in familial Alzheimer's disease and Huntington's disease.}}
\newblock \bibinfo{journal}{NeuroImage} \bibinfo{volume}{60},
  \bibinfo{pages}{1880--1889}.
\bibitem[{Galecki and Burzykowski(2013)}]{Galecki2013-ch10}
\bibinfo{author}{Galecki, A.}, \bibinfo{author}{Burzykowski, T.},
  \bibinfo{year}{2013}.
\newblock \bibinfo{title}{{Linear model with fixed effects and correlated
  errors}}, in: \bibinfo{booktitle}{Linear Mixed-Effects Models Using R}.
  \bibinfo{publisher}{Springer New York}, \bibinfo{address}{New York, NY}.
  Springer Texts in Statistics. chapter~\bibinfo{chapter}{10}, pp.
  \bibinfo{pages}{177--196}.
\bibitem[{Ito et~al.(2011)Ito, Corrigan, Zhao, French, Miller, Soares, Katz,
  Nicholas, Billing, Anziano and Fullerton}]{Ito2011}
\bibinfo{author}{Ito, K.}, \bibinfo{author}{Corrigan, B.},
  \bibinfo{author}{Zhao, Q.}, \bibinfo{author}{French, J.},
  \bibinfo{author}{Miller, R.}, \bibinfo{author}{Soares, H.},
  \bibinfo{author}{Katz, E.}, \bibinfo{author}{Nicholas, T.},
  \bibinfo{author}{Billing, B.}, \bibinfo{author}{Anziano, R.},
  \bibinfo{author}{Fullerton, T.}, \bibinfo{year}{2011}.
\newblock \bibinfo{title}{{Disease progression model for cognitive
  deterioration from Alzheimer's Disease Neuroimaging Initiative database}}.
\newblock \bibinfo{journal}{Alzheimer's and Dementia} \bibinfo{volume}{7},
  \bibinfo{pages}{151--160}.
\bibitem[{Jack et~al.(2013)Jack, Knopman, Jagust, Petersen, Weiner, Aisen,
  Shaw, Vemuri, Wiste, Weigand, Lesnick, Pankratz, Donohue and
  Trojanowski}]{Jack2013}
\bibinfo{author}{Jack, C.R.}, \bibinfo{author}{Knopman, D.S.},
  \bibinfo{author}{Jagust, W.J.}, \bibinfo{author}{Petersen, R.C.},
  \bibinfo{author}{Weiner, M.W.}, \bibinfo{author}{Aisen, P.S.},
  \bibinfo{author}{Shaw, L.M.}, \bibinfo{author}{Vemuri, P.},
  \bibinfo{author}{Wiste, H.J.}, \bibinfo{author}{Weigand, S.D.},
  \bibinfo{author}{Lesnick, T.G.}, \bibinfo{author}{Pankratz, V.S.},
  \bibinfo{author}{Donohue, M.C.}, \bibinfo{author}{Trojanowski, J.Q.},
  \bibinfo{year}{2013}.
\newblock \bibinfo{title}{{Tracking pathophysiological processes in Alzheimer's
  disease: an updated hypothetical model of dynamic biomarkers}}.
\newblock \bibinfo{journal}{Lancet Neurology} \bibinfo{volume}{12},
  \bibinfo{pages}{207--216}.
\bibitem[{Jack et~al.(2008)Jack, Lowe, Senjem, Weigand, Kemp, Shiung, Knopman,
  Boeve, Klunk, Mathis and Petersen}]{Jack2008}
\bibinfo{author}{Jack, C.R.}, \bibinfo{author}{Lowe, V.J.},
  \bibinfo{author}{Senjem, M.L.}, \bibinfo{author}{Weigand, S.D.},
  \bibinfo{author}{Kemp, B.J.}, \bibinfo{author}{Shiung, M.M.},
  \bibinfo{author}{Knopman, D.S.}, \bibinfo{author}{Boeve, B.F.},
  \bibinfo{author}{Klunk, W.E.}, \bibinfo{author}{Mathis, C.a.},
  \bibinfo{author}{Petersen, R.C.}, \bibinfo{year}{2008}.
\newblock \bibinfo{title}{{$^{11}$C PiB and structural MRI provide
  complementary information in imaging of Alzheimer's disease and amnestic mild
  cognitive impairment}}.
\newblock \bibinfo{journal}{Brain} \bibinfo{volume}{131},
  \bibinfo{pages}{665--680}.
\bibitem[{Jedynak et~al.(2012)Jedynak, Lang, Liu, Katz, Zhang, Wyman, Raunig,
  Jedynak, Caffo and Prince}]{Jedynak2012}
\bibinfo{author}{Jedynak, B.M.}, \bibinfo{author}{Lang, A.},
  \bibinfo{author}{Liu, B.}, \bibinfo{author}{Katz, E.},
  \bibinfo{author}{Zhang, Y.}, \bibinfo{author}{Wyman, B.T.},
  \bibinfo{author}{Raunig, D.}, \bibinfo{author}{Jedynak, C.P.},
  \bibinfo{author}{Caffo, B.}, \bibinfo{author}{Prince, J.L.},
  \bibinfo{year}{2012}.
\newblock \bibinfo{title}{{A computational neurodegenerative disease
  progression score: Method and results with the Alzheimer's disease
  Neuroimaging Initiative cohort}}.
\newblock \bibinfo{journal}{NeuroImage} \bibinfo{volume}{63},
  \bibinfo{pages}{1478--1486}.
\bibitem[{Jedynak et~al.(2014)Jedynak, Liu, Lang, Gel and Prince}]{Jedynak2014}
\bibinfo{author}{Jedynak, B.M.}, \bibinfo{author}{Liu, B.},
  \bibinfo{author}{Lang, A.}, \bibinfo{author}{Gel, Y.},
  \bibinfo{author}{Prince, J.L.}, \bibinfo{year}{2014}.
\newblock \bibinfo{title}{{A computational method for computing an Alzheimer's
  disease progression score; experiments and validation with the ADNI data
  set.}}
\newblock \bibinfo{journal}{Neurobiology of Aging} \bibinfo{volume}{36
  Supplement}, \bibinfo{pages}{S178--S184}.
\bibitem[{Jenkinson et~al.(2002)Jenkinson, Bannister, Brady and
  Smith}]{Jenkinson2002}
\bibinfo{author}{Jenkinson, M.}, \bibinfo{author}{Bannister, P.},
  \bibinfo{author}{Brady, M.}, \bibinfo{author}{Smith, S.},
  \bibinfo{year}{2002}.
\newblock \bibinfo{title}{{Improved optimization for the robust and accurate
  linear registration and motion correction of brain images}}.
\newblock \bibinfo{journal}{NeuroImage} \bibinfo{volume}{17},
  \bibinfo{pages}{825--841}.
\bibitem[{Lindstrom and Bates(1990)}]{Lindstrom1990}
\bibinfo{author}{Lindstrom, M.}, \bibinfo{author}{Bates, D.},
  \bibinfo{year}{1990}.
\newblock \bibinfo{title}{{Nonlinear mixed effects models for repeated measures
  data}}.
\newblock \bibinfo{journal}{Biometrics} \bibinfo{volume}{46},
  \bibinfo{pages}{673--687}.
\bibitem[{Mintun et~al.(2006)Mintun, Larossa, Sheline, Dence, Lee, Mach, Klunk,
  Mathis, DeKosky and Morris}]{Mintun2006}
\bibinfo{author}{Mintun, M.A.}, \bibinfo{author}{Larossa, G.N.},
  \bibinfo{author}{Sheline, Y.I.}, \bibinfo{author}{Dence, C.S.},
  \bibinfo{author}{Lee, S.Y.}, \bibinfo{author}{Mach, R.H.},
  \bibinfo{author}{Klunk, W.E.}, \bibinfo{author}{Mathis, C.A.},
  \bibinfo{author}{DeKosky, S.T.}, \bibinfo{author}{Morris, J.C.},
  \bibinfo{year}{2006}.
\newblock \bibinfo{title}{{[¹¹C]PIB in a nondemented population: potential
  antecedent marker of Alzheimer disease}}.
\newblock \bibinfo{journal}{Neurology} \bibinfo{volume}{67},
  \bibinfo{pages}{446--452}.
\bibitem[{Pinheiro and Bates(1996)}]{Pinheiro1996}
\bibinfo{author}{Pinheiro, J.C.}, \bibinfo{author}{Bates, D.M.},
  \bibinfo{year}{1996}.
\newblock \bibinfo{title}{{Unconstrained parametrizations for
  variance-covariance matrices}}.
\newblock \bibinfo{journal}{Statistics and Computing} \bibinfo{volume}{6},
  \bibinfo{pages}{289--296}.
\bibitem[{Resnick et~al.(2000)Resnick, Goldszal, Davatzikos, Golski, Kraut,
  Metter, Bryan and Zonderman}]{Resnick2000}
\bibinfo{author}{Resnick, S.M.}, \bibinfo{author}{Goldszal, A.F.},
  \bibinfo{author}{Davatzikos, C.}, \bibinfo{author}{Golski, S.},
  \bibinfo{author}{Kraut, M.A.}, \bibinfo{author}{Metter, E.J.},
  \bibinfo{author}{Bryan, R.N.}, \bibinfo{author}{Zonderman, A.B.},
  \bibinfo{year}{2000}.
\newblock \bibinfo{title}{{One-year age changes in MRI brain volumes in older
  adults}}.
\newblock \bibinfo{journal}{Cerebral Cortex} \bibinfo{volume}{10},
  \bibinfo{pages}{464--472}.
\bibitem[{Rodrigue et~al.(2012)Rodrigue, Kennedy, {Devous Sr.}, Rieck, Hebrank,
  Diaz-Arrastia, Mathews and Park}]{Rodrigue2012}
\bibinfo{author}{Rodrigue, K.M.}, \bibinfo{author}{Kennedy, K.M.},
  \bibinfo{author}{{Devous Sr.}, M.D.}, \bibinfo{author}{Rieck, J.R.},
  \bibinfo{author}{Hebrank, A.C.}, \bibinfo{author}{Diaz-Arrastia, R.},
  \bibinfo{author}{Mathews, D.}, \bibinfo{author}{Park, D.C.},
  \bibinfo{year}{2012}.
\newblock \bibinfo{title}{$\beta$-amyloid burden in healthy aging}.
\newblock \bibinfo{journal}{Neurology} \bibinfo{volume}{78},
  \bibinfo{pages}{387--395}.
\bibitem[{Schiratti et~al.(2015a)Schiratti, Allassonniere, Colliot and
  Durrleman}]{Schiratti2015a}
\bibinfo{author}{Schiratti, J.B.}, \bibinfo{author}{Allassonniere, S.},
  \bibinfo{author}{Colliot, O.}, \bibinfo{author}{Durrleman, S.},
  \bibinfo{year}{2015}a.
\newblock \bibinfo{title}{{Learning spatiotemporal trajectories from
  manifold-valued longitudinal data}}, in: \bibinfo{editor}{Cortes, C.},
  \bibinfo{editor}{Lawrence, N.}, \bibinfo{editor}{Lee, D.},
  \bibinfo{editor}{Sugiyama, M.}, \bibinfo{editor}{Garnett, R.} (Eds.),
  \bibinfo{booktitle}{Advances in Neural Information Processing Systems 28},
  pp. \bibinfo{pages}{2395--2403}.
\bibitem[{Schiratti et~al.(2015b)Schiratti, Allassonniere, Routier, Colliot and
  Durrleman}]{Schiratti2015}
\bibinfo{author}{Schiratti, J.B.}, \bibinfo{author}{Allassonniere, S.},
  \bibinfo{author}{Routier, A.}, \bibinfo{author}{Colliot, O.},
  \bibinfo{author}{Durrleman, S.}, \bibinfo{year}{2015}b.
\newblock \bibinfo{title}{{A mixed-effects model with time reparametrization
  for longitudinal univariate manifold-valued data}}, in:
  \bibinfo{editor}{Ourselin, S.}, \bibinfo{editor}{Alexander, D.C.},
  \bibinfo{editor}{Westin, C.F.}, \bibinfo{editor}{Cardoso, M.J.} (Eds.),
  \bibinfo{booktitle}{Lecture Notes in Computer Science (9123), Information
  Processing in Medical Imaging}, pp. \bibinfo{pages}{564--575}.
\bibitem[{Schmidt-Richberg et~al.(2015)Schmidt-Richberg, Guerrero, Ledig,
  Molina-Abril, Frangi and Rueckert}]{Schmidt-Richberg2015}
\bibinfo{author}{Schmidt-Richberg, A.}, \bibinfo{author}{Guerrero, R.},
  \bibinfo{author}{Ledig, C.}, \bibinfo{author}{Molina-Abril, H.},
  \bibinfo{author}{Frangi, A.F.}, \bibinfo{author}{Rueckert, D.},
  \bibinfo{year}{2015}.
\newblock \bibinfo{title}{{Multi-stage biomarker models for progression
  estimation in Alzheimer's disease}}.
\newblock \bibinfo{journal}{Lecture Notes in Computer Science (9123),
  Information Processing in Medical Imaging} \bibinfo{volume}{9123},
  \bibinfo{pages}{387--398}.
\bibitem[{Schulam et~al.(2015)Schulam, Wigley and Saria}]{Schulam2015}
\bibinfo{author}{Schulam, P.}, \bibinfo{author}{Wigley, F.},
  \bibinfo{author}{Saria, S.}, \bibinfo{year}{2015}.
\newblock \bibinfo{title}{{Clustering longitudinal clinical marker trajectories
  from electronic health data: Applications to phenotyping and endotype
  discovery}}, in: \bibinfo{booktitle}{AAAI Conference on Artificial
  Intelligence}.
\bibitem[{Shock et~al.(1984)Shock, Greulich, Andres, Arenberg, {Costa Jr.},
  Lakatta and Tobin}]{Shock1984}
\bibinfo{author}{Shock, N.W.}, \bibinfo{author}{Greulich, R.C.},
  \bibinfo{author}{Andres, R.}, \bibinfo{author}{Arenberg, D.},
  \bibinfo{author}{{Costa Jr.}, P.T.}, \bibinfo{author}{Lakatta, E.G.},
  \bibinfo{author}{Tobin, J.D.}, \bibinfo{year}{1984}.
\newblock \bibinfo{title}{{Normal human aging: The Baltimore Longitudinal Study
  of Aging}}.
\newblock \bibinfo{type}{Technical Report}. U.S. Government Printing Office.
  \bibinfo{address}{Washington, DC}.
\bibitem[{Sperling et~al.(2014a)Sperling, Mormino and Johnson}]{Sperling2014b}
\bibinfo{author}{Sperling, R.}, \bibinfo{author}{Mormino, E.},
  \bibinfo{author}{Johnson, K.}, \bibinfo{year}{2014}a.
\newblock \bibinfo{title}{{The evolution of preclinical Alzheimer's disease:
  implications for prevention trials}}.
\newblock \bibinfo{journal}{Neuron} \bibinfo{volume}{84},
  \bibinfo{pages}{608--622}.
\bibitem[{Sperling et~al.(2014b)Sperling, Rentz, Johnson, Karlawish, Donohue,
  Salmon and Aisen}]{Sperling2014}
\bibinfo{author}{Sperling, R.A.}, \bibinfo{author}{Rentz, D.M.},
  \bibinfo{author}{Johnson, K.A.}, \bibinfo{author}{Karlawish, J.},
  \bibinfo{author}{Donohue, M.}, \bibinfo{author}{Salmon, D.P.},
  \bibinfo{author}{Aisen, P.}, \bibinfo{year}{2014}b.
\newblock \bibinfo{title}{{The A4 Study: Stopping AD Before Symptoms Begin?}}
\newblock \bibinfo{journal}{Science Translational Medicine}
  \bibinfo{volume}{6}, \bibinfo{pages}{228fs13}.
\bibitem[{Villain et~al.(2012)Villain, Ch{\'{e}}telat, Grassiot, Bourgeat,
  Jones, Ellis, Ames, Martins, Eustache, Salvado, Masters, Rowe and
  Villemagne}]{Villain2012}
\bibinfo{author}{Villain, N.}, \bibinfo{author}{Ch{\'{e}}telat, G.},
  \bibinfo{author}{Grassiot, B.}, \bibinfo{author}{Bourgeat, P.},
  \bibinfo{author}{Jones, G.}, \bibinfo{author}{Ellis, K.A.},
  \bibinfo{author}{Ames, D.}, \bibinfo{author}{Martins, R.N.},
  \bibinfo{author}{Eustache, F.}, \bibinfo{author}{Salvado, O.},
  \bibinfo{author}{Masters, C.L.}, \bibinfo{author}{Rowe, C.C.},
  \bibinfo{author}{Villemagne, V.L.}, \bibinfo{year}{2012}.
\newblock \bibinfo{title}{{Regional dynamics of amyloid-$\beta$ deposition in
  healthy elderly, mild cognitive impairment and Alzheimer's disease: a
  voxelwise PiB-PET longitudinal study}}.
\newblock \bibinfo{journal}{Brain} \bibinfo{volume}{135},
  \bibinfo{pages}{2126--2139}.
\bibitem[{Villemagne et~al.(2013)Villemagne, Burnham, Bourgeat, Brown, Ellis,
  Salvado, Szoeke, Macaulay, Martins, Maruff, Ames, Rowe and
  Masters}]{Villemagne2013}
\bibinfo{author}{Villemagne, V.L.}, \bibinfo{author}{Burnham, S.},
  \bibinfo{author}{Bourgeat, P.}, \bibinfo{author}{Brown, B.},
  \bibinfo{author}{Ellis, K.A.}, \bibinfo{author}{Salvado, O.},
  \bibinfo{author}{Szoeke, C.}, \bibinfo{author}{Macaulay, S.L.},
  \bibinfo{author}{Martins, R.}, \bibinfo{author}{Maruff, P.},
  \bibinfo{author}{Ames, D.}, \bibinfo{author}{Rowe, C.C.},
  \bibinfo{author}{Masters, C.L.}, \bibinfo{year}{2013}.
\newblock \bibinfo{title}{{Amyloid $\beta$ deposition, neurodegeneration, and
  cognitive decline in sporadic Alzheimer's disease: a prospective cohort
  study}}.
\newblock \bibinfo{journal}{Lancet Neurology} \bibinfo{volume}{12},
  \bibinfo{pages}{357--367}.
\bibitem[{Villeneuve et~al.(2015)Villeneuve, Rabinovici, Cohn-Sheehy, Madison,
  Ayakta, Ghosh, {La Joie}, Arthur-Bentil, Vogel, Marks, Lehmann, Rosen, Reed,
  Olichney, Boxer, Miller, Borys, Jin, Huang, Grinberg, DeCarli, Seeley and
  Jagust}]{Villeneuve2015}
\bibinfo{author}{Villeneuve, S.}, \bibinfo{author}{Rabinovici, G.D.},
  \bibinfo{author}{Cohn-Sheehy, B.I.}, \bibinfo{author}{Madison, C.},
  \bibinfo{author}{Ayakta, N.}, \bibinfo{author}{Ghosh, P.M.},
  \bibinfo{author}{{La Joie}, R.}, \bibinfo{author}{Arthur-Bentil, S.K.},
  \bibinfo{author}{Vogel, J.W.}, \bibinfo{author}{Marks, S.M.},
  \bibinfo{author}{Lehmann, M.}, \bibinfo{author}{Rosen, H.J.},
  \bibinfo{author}{Reed, B.}, \bibinfo{author}{Olichney, J.},
  \bibinfo{author}{Boxer, A.L.}, \bibinfo{author}{Miller, B.L.},
  \bibinfo{author}{Borys, E.}, \bibinfo{author}{Jin, L.W.},
  \bibinfo{author}{Huang, E.J.}, \bibinfo{author}{Grinberg, L.T.},
  \bibinfo{author}{DeCarli, C.}, \bibinfo{author}{Seeley, W.W.},
  \bibinfo{author}{Jagust, W.}, \bibinfo{year}{2015}.
\newblock \bibinfo{title}{{Existing Pittsburgh Compound-B positron emission
  tomography thresholds are too high: statistical and pathological
  evaluation}}.
\newblock \bibinfo{journal}{Brain} \bibinfo{volume}{138},
  \bibinfo{pages}{2020--2033}.
\bibitem[{Yang et~al.(2011)Yang, Farnum, Lobanov, Schultz, Raghavan, Samtani,
  Novak, Narayan and DiBernardo}]{Yang2011}
\bibinfo{author}{Yang, E.}, \bibinfo{author}{Farnum, M.},
  \bibinfo{author}{Lobanov, V.}, \bibinfo{author}{Schultz, T.},
  \bibinfo{author}{Raghavan, N.}, \bibinfo{author}{Samtani, M.N.},
  \bibinfo{author}{Novak, G.}, \bibinfo{author}{Narayan, V.},
  \bibinfo{author}{DiBernardo, A.}, \bibinfo{year}{2011}.
\newblock \bibinfo{title}{{Quantifying the pathophysiological timeline of
  Alzheimer's disease}}.
\newblock \bibinfo{journal}{Journal of Alzheimer's Disease}
  \bibinfo{volume}{26}, \bibinfo{pages}{745--753}.
\bibitem[{Younes et~al.(2014)Younes, Albert and Miller}]{Younes2014}
\bibinfo{author}{Younes, L.}, \bibinfo{author}{Albert, M.},
  \bibinfo{author}{Miller, M.I.}, \bibinfo{year}{2014}.
\newblock \bibinfo{title}{{Inferring changepoint times of medial temporal lobe
  morphometric change in preclinical Alzheimer's disease.}}
\newblock \bibinfo{journal}{NeuroImage: Clinical} \bibinfo{volume}{5},
  \bibinfo{pages}{178--187}.
\bibitem[{Young et~al.(2014)Young, Oxtoby, Daga, Cash, Fox, Ourselin, Schott
  and Alexander}]{Young2014}
\bibinfo{author}{Young, A.L.}, \bibinfo{author}{Oxtoby, N.P.},
  \bibinfo{author}{Daga, P.}, \bibinfo{author}{Cash, D.M.},
  \bibinfo{author}{Fox, N.C.}, \bibinfo{author}{Ourselin, S.},
  \bibinfo{author}{Schott, J.M.}, \bibinfo{author}{Alexander, D.C.},
  \bibinfo{year}{2014}.
\newblock \bibinfo{title}{{A data-driven model of biomarker changes in sporadic
  Alzheimer's disease}}.
\newblock \bibinfo{journal}{Brain} \bibinfo{volume}{137},
  \bibinfo{pages}{2564--2577}.
\bibitem[{Zhou et~al.(2003)Zhou, Endres, Bra\v{s}i\'{c}, Huang and
  Wong}]{Zhou2003}
\bibinfo{author}{Zhou, Y.}, \bibinfo{author}{Endres, C.J.},
  \bibinfo{author}{Bra\v{s}i\'{c}, J.R.}, \bibinfo{author}{Huang, S.C.},
  \bibinfo{author}{Wong, D.F.}, \bibinfo{year}{2003}.
\newblock \bibinfo{title}{{Linear regression with spatial constraint to
  generate parametric images of ligand-receptor dynamic PET studies with a
  simplified reference tissue model}}.
\newblock \bibinfo{journal}{NeuroImage} \bibinfo{volume}{18},
  \bibinfo{pages}{975--989}.
\bibitem[{Ziegler et~al.(2015)Ziegler, Penny, Ridgway, Ourselin and
  Friston}]{Ziegler2015}
\bibinfo{author}{Ziegler, G.}, \bibinfo{author}{Penny, W.D.},
  \bibinfo{author}{Ridgway, G.R.}, \bibinfo{author}{Ourselin, S.},
  \bibinfo{author}{Friston, K.J.}, \bibinfo{year}{2015}.
\newblock \bibinfo{title}{{Estimating anatomical trajectories with Bayesian
  mixed-effects modeling}}.
\newblock \bibinfo{journal}{NeuroImage} \bibinfo{volume}{121},
  \bibinfo{pages}{51--68}.

\end{thebibliography}

\end{document}